\documentclass[journal]{IEEEtran}
\IEEEoverridecommandlockouts
\IEEEpubid{0000--0000/00\$00.00~\copyright~2023 IEEE}
\usepackage{epsfig}
\usepackage{graphicx}           
\usepackage{amssymb}
\usepackage{amsmath}
\usepackage{upgreek}
\usepackage{subcaption}
\usepackage{bm}
\usepackage{overpic}
\usepackage{float}
\usepackage{threeparttable}
\usepackage{tikz}
\usepackage{tabularx}
\newlength{\he}
\AtBeginDocument{\settoheight{\he}{A}}
\usepackage[T1]{fontenc} % with this, the abstract is 'bold'
\usepackage{multirow} 
\usepackage{tablefootnote}
\usepackage[normalem]{ulem}
\usepackage{cite}
\newcommand{\red}{\color{red}}

\title{A 3.5 GS/s 1-1 MASH VCO ADC With Second-order Noise Shaping}
\author{Brendan Saux, Jonas Borgmans,~\IEEEmembership{Member,~IEEE,} Johan Raman and Pieter Rombouts,~\IEEEmembership{Senior~Member,~IEEE}
        \thanks{The authors are with the Department of Electronics and Information Systems (ELIS), Ghent University, 9052 Gent, Belgium (e-mail: brendan.saux@ugent.be).}}

\begin{document}

\maketitle
\begin{abstract}
   In this work, a 3.5 GS/s voltage-controlled oscillator (VCO) analog-to-digital converter (ADC) using multi-stage noise shaping (MASH) is presented. This 28nm CMOS ADC achieves second-order noise shaping in an easily-scalable, open-loop  configuration. A key enabler of the high-bandwidth MASH VCO ADC is the use of a multi-bit estimated error signal. With an OSR of 16, an SNDR of 67 dB and DR of 68 dB are achieved in 109.375 MHz bandwidth. The full-custom pseudo-analog circuits consume 9 mW, while the automatically generated digital circuits consume another 24 mW. A $\mathbf{FoM_{DR} = 163}$ dB and core area of $\mathbf{0.017\,\mathbf{mm}^2}$ are obtained.
\end{abstract}

\begin{IEEEkeywords} analog-to-digital converter (ADC), voltage-controlled oscillator (VCO), multi-stage noise shaping (MASH), second-order, scalable
\end{IEEEkeywords}
    
\section{Introduction\label{sect:intro}}

Over the past few years, ring oscillator-based voltage-controlled oscillator (VCO) analog-to-digital converters (ADCs) have gained popularity~\cite{ gielen2020time2, bookVCOADC}.
They offer a multitude of interesting properties, such as inherent anti-aliasing, guaranteed monotonicity,
a very low intrinsic noise level and a high sensitivity. Even more enticing is their ability to leverage CMOS technology scaling without the challenges that traditional analog circuits face. This has led to their adoption in a wide range of fields, including sensor readouts, IoT and biomedical applications 
\cite{nguyen2021deep,baert2020,shibata202016,wu2019,amir_2017,wu202016,xing201542,aiello2021capacitance,park_perrott,reddy201216,16nm_huang,dey201750,jiang20202,sacco202016,mukherjee202074,huang2020112,watanabe2021,pochet202128}.

An important next step to improve the performance  of VCO A/D conversion is increasing the noise shaping order at high bandwidths ($>$100 MHz). 
Multiple higher-order VCO ADC architectures have already been proposed and implemented. However, these options are often not very well suited to our goal of creating an easily scalable and high speed ADC. 

For instance, designs as \cite{Temes2020TCAS1, reddy201216, dey201750, 16nm_huang, hanumolu_3order, park_perrott, zhu2015, guo_2022} reach impressive performance. However, next to VCOs, they also depend on loop filters using OTAs and are therefore not as easily scalable, which counters one of the principal advantages of VCO-based A/D conversion. Higher-order continuous-time $\Sigma \Delta$-modulators using only VCOs have also been presented \cite{amir_2017, Cardes2018JSSC, Sanyal2019, zhong_2020, li_2020}, but these can be challenging to design for high bandwidths due to the trade-off between performance and stability linked to excess loop delay. 

We therefore believe 1-1 multi-stage noise shaping (MASH) VCO ADCs to be a promising option for high bandwidths, as they enable purely VCO-based  higher-order A/D conversion without this trade-off. Although this technique has been implemented for bandwidths of up to 2 MHz \cite{MaghamiJSSC2020}, to date, no successful designs using this technique at high bandwidths have been presented. The main difficulty is realizing a sufficiently accurate coupling between the successive stages. 

In this work, the coupling between the two stages is performed by using a multi-bit estimated error signal, which allows multi-phase readout of the first stage. The potential issue of nonlinearity in the second stage, which leads to noise leakage, is solved by the use of a pseudo-differential setup using cross-coupling. We will show this method also offers an increased robustness against pulse-width errors. Special care was taken to maintain the integrity of the estimated error signal by the use of dedicated buffers to pull the VCO output signal rail-to-rail and the use of fast sense amplifiers in the error estimation circuit to reduce metastability.

This results in the first implementation of a higher-order purely VCO-based ADC with a bandwidth greater than 100\ MHz. This is also the first implementation of a 1-1 MASH VCO ADC featuring multi-bit error estimation. 

The rest of this paper is organised as follows. In section \ref{sect:sys} the implemented architecture using the multi-bit estimated error signal is presented and analyzed at the system-level. In section \ref{sect:cross} the issue of nonlinearity in the second VCO ADC stage is discussed along with its system-level solution. Section \ref{sect:cir} details the circuit-level implementation and design considerations. Section \ref{sect:meas} contains measurement results and explains the calibration procedure to suppress harmonics. Finally, section \ref{sect:conc} concludes this work.
\IEEEpubidadjcol
\section{MASH VCO A/D Conversion \label{sect:sys}} 
\begin{figure}
    \center
    \includegraphics[width=\columnwidth]{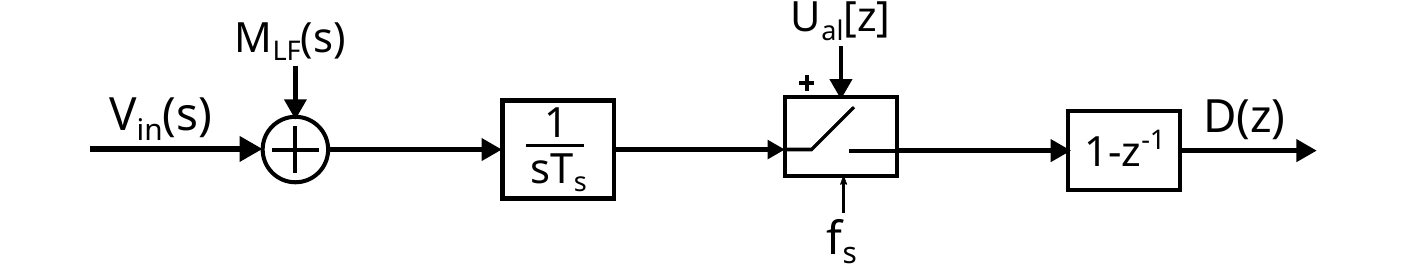}
    \caption{Simplified PFM model of a single-stage VCO ADC showing in-band signals.} \label{fig:sys_vco}
\end{figure}
\begin{figure}
    \center
    \includegraphics[width=\columnwidth]{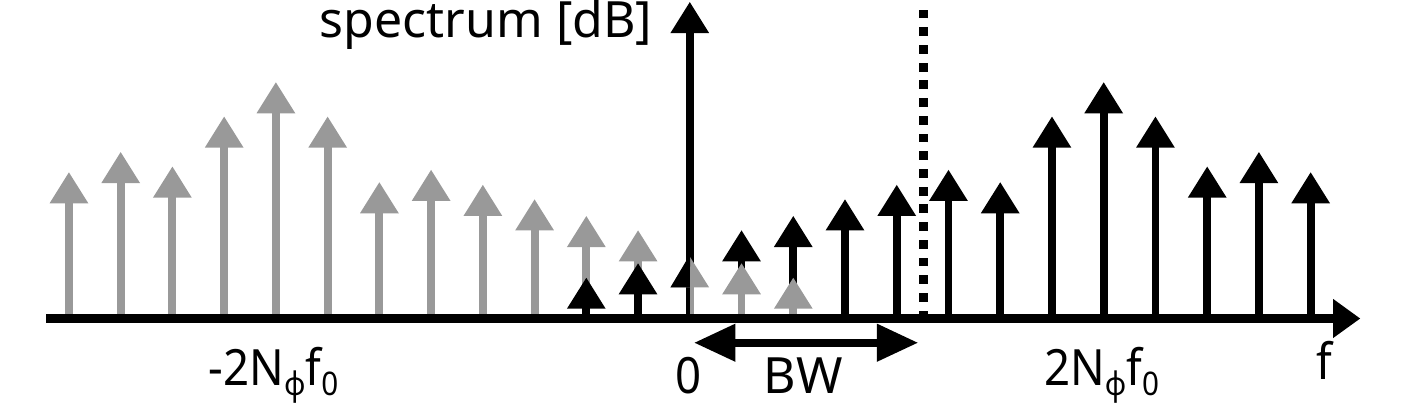}
    \caption{Conceptual illustration of the performance degradation due to presence of the spurs of the first PFM sideband in the signal bandwidth \cite{gutierrez2017pulse}.} \label{fig:sys_pfm}
\end{figure}
\begin{figure*}[t]
    \center
    \includegraphics[width=\textwidth]{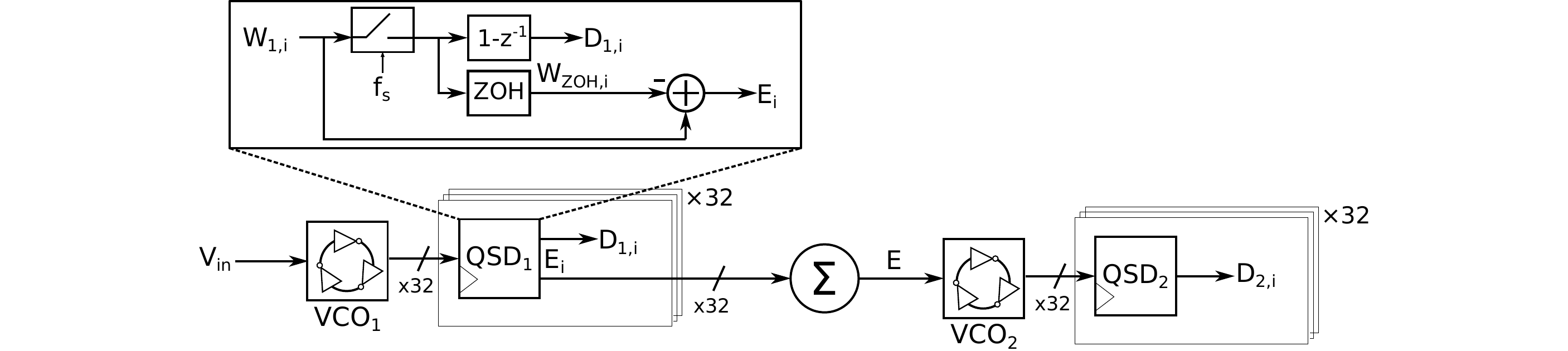}
    \caption{Initial MASH VCO architecture. }
    \label{fig:system_naive}
\end{figure*}
In the following sections, explicit references to time and frequency variables in the text will be omitted for readability, e.g. $E$ will be used instead of $E(s)$. 
\subsection{Multi-Bit Error Estimation}
This work builds on \cite{gutierrez2017pulse}, where it was shown that a single-stage VCO ADC is strictly equivalent to a pulse frequency modulator (PFM) followed by a pulse shaping filter and a sampler. A few of the insights obtained from this model will be essential to the understanding of the implemented architecture. {We will summerize these here first.}

A model of a simple VCO ADC with relevant in-band signals is shown in Fig.~\ref{fig:sys_vco}. For simplicity, the VCO frequency gain $K$ was normalized to 1 and the contribution of the VCO rest frequency $f_0$ was omitted. An input signal $V_{in}$ is applied to the system and is subsequently integrated, sampled and differentiated to obtain the output signal $D$. Two noise sources are present in the system: $M_{LF}$ represents the in-band PFM components and $U_{al}$ represents the aliased high-frequency PFM components that enter the bandwidth through sampling. $U_{al}$ is what is commonly referred to as the quantization noise in the phase model interpretation of VCO A/D conversion. However, $M_{LF}$, which consists of spurious tones of the first PFM sideband, also degrades the performance of the system. Assuming the readout circuits use both the rising and falling edges of the VCO, the first PFM sideband is centered around the effective VCO rest frequency $f_{eff} = 2 N_\phi f_0$ as shown in Fig.~\ref{fig:sys_pfm}. Here $N_\phi$ represents the number of readout phases.

Two insights are now paramount. The first is that $M_{LF}$ cannot be removed from the system and therefore represents a fundamental limit \cite{gutierrez2017pulse}. The only way to reduce it for a given bandwidth is either by increasing $f_0$ - which is only possible until a certain point - or boosting the number of readout phases $N_\phi$. It is therefore desirable to use a high $N_\phi$, especially at high bandwidths.

The second is that the `quantization noise' $U_{al}$ is added for each VCO phase \textit{at the moment of sampling}. Therefore, by subtracting a VCO output and its sampled and held value, the quantization noise added at the moment of sampling of that phase can be estimated. The total quantization noise in this VCO ADC stage, which is the sum of the quantization noise of each phase, can then be estimated by summing the estimated error signals of each phase. This multi-bit global error can be fed to the input of a second ADC stage. By combining the output of the two stages using the correct noise cancellation filters (NCF), $U_{al}$ of the first stage can be cancelled. This enables the efficient implementation of 1-1 MASH architectures using multi-phase readout in the first stage.

In contrast, almost all previously published VCO-based 1-1 MASH \cite{hernandez_mash, yu_vco} or sturdy MASH \cite{sacco202016} ADC architectures use a single-phase readout in the first stage. These architectures will therefore be limited to low bandwidths to limit the effect of the low-frequency PFM spurs $M_{LF}$, since the first PFM sideband will be located at the much lower frequency of $2f_0$. Since this is solved in our architecture, we expect a substantially higher performance at high bandwidths. This is discussed in more detail after a description and analysis of our architecture.
\begin{figure}
    \centering
    \includegraphics[width=1\columnwidth]{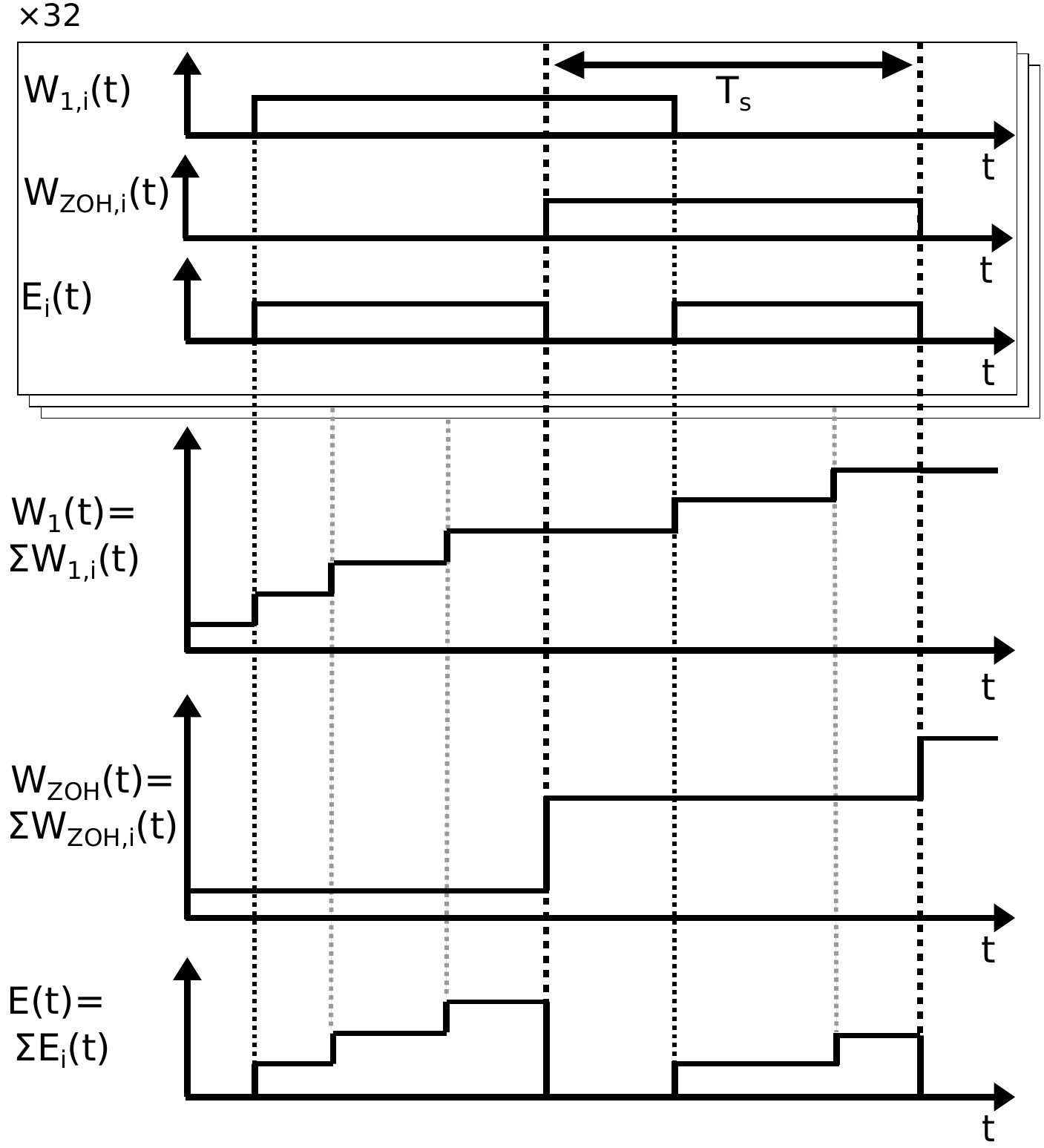}   
    \caption{Illustration of time-domain waveforms of the relevant signals to obtain the estimated error signal $E$. }\label{fig:multibit}
\end{figure}
\subsection{Initial 1-1 MASH VCO Architecture}
Fig.~\ref{fig:system_naive} shows a first iteration of the implemented 1-1 MASH VCO architecture. In this, a 32-phase VCO readout is used, where each phase $W_i$ of the first stage is followed by a readout quantization/sampling/differentiation (QSD) block \cite{borgmans2021methodology}. These QSDs generate the VCO phases' quantized 1-bit outputs $D_{1,i}$, which are afterwards combined to form the overall output $D_1$ of the first stage. The QSDs are modified to also perform the error estimation (see section \ref{subsect:qsd} for the circuit implementation) and generate 1-bit pulse signals $E_{i}$ which represent the quantization error. The 1-bit pulse signals each contain the quantization error of their respective VCO phase and can therefore be summed to obtain a global estimated error signal $E$ as explained above. Fig.~\ref{fig:multibit} shows the relevant time-domain waveforms to obtain $E$, which is identical to what would be obtained using a multi-bit counter that counts the individual phase edges. {Notice that the information in $E_i$ is contained in the individual pulse lengths.}

Finally, each phase of the second stage is also followed by a QSD, which generates the 1-bit outputs $D_{2,i}$. These are combined to get the overall output $D_2$ of the second stage.
In this initial setup, a rather straightforward way of driving the second stage is used. The estimated error signals $E_{i}$ are immediately summed to calculate $E$, which is fed to the second stage.

\subsection{Analysis and Theoretical Performance \label{subsect:analysis}}
We will now perform an analysis of the architecture of Fig.~\ref{fig:system_naive}. To provide a more easily accessible discussion of this architecture, it is performed here using the more conventional phase model interpretation of VCO A/D conversion instead of the PFM model discussed above. However, it must be stressed that this is only valid if the number of readout phases in the first and second stage are high enough to be able to neglect the in-band PFM spurs $M_{LF}$.

\begin{figure}[t]
    \center
    \includegraphics[width=\columnwidth]{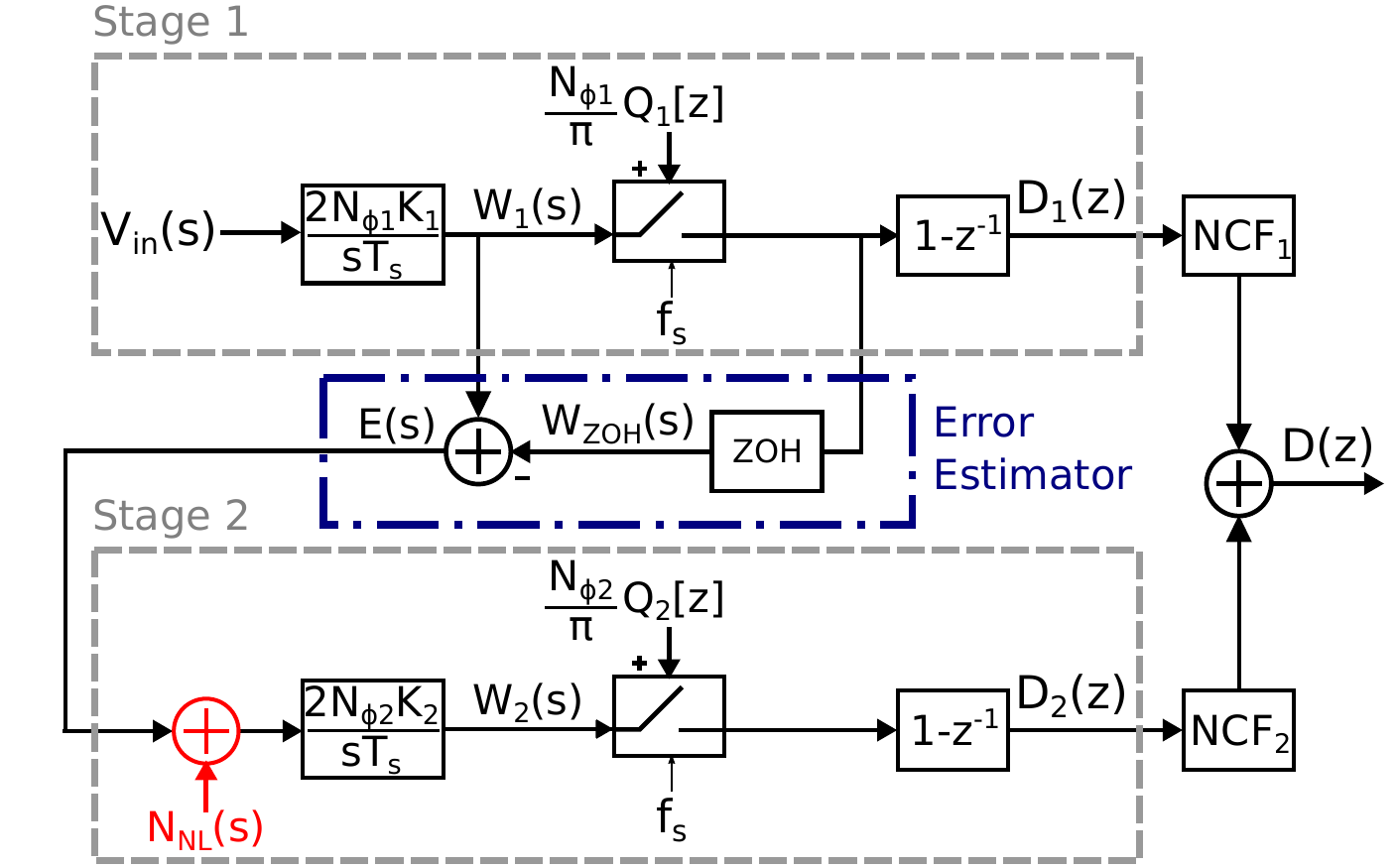}
    \caption{Simplified block diagram of a MASH VCO ADC (black). The equivalent error $N_{NL}$ to model the effect of nonlinearity in the second stage (red) was also added.}\label{fig:sys_mash}
\end{figure}

A block diagram representation of the 1-1 MASH architecture using the phase model is shown in Fig.~\ref{fig:sys_mash} (black). The free-running frequency $f_0$ was ignored for clarity. $N_{\phi 1}$, $K_{1}$ and $Q_1$ respectively represent the number of phases, frequency gain and {phase} quantization error of the first stage. $N_{\phi 2}$, $K_{2}$ and $Q_2$ represent the same concepts for the second stage.

The output of the first stage can be written as in \cite{vco_kim_jang} by
\begin{equation}
    D_1(z)= \frac{2N_{\phi 1} K_1}{f_s}\cdot[V_{in}(s)]^*+\frac{N_{\phi 1}}{\pi}\cdot Q_1(z)\cdot(1-z^{-1})
\label{eq:y1vco}
\end{equation}
where the star operator $[\cdot]^*$ indicates the effect of sampling \cite{tuxal1955}. For simplicity, the $\mathrm{sinc}(sT_s)$ leading to anti-aliasing was assumed to be approximately equal to 1 in the signal band.

The two stages are coupled through the error estimator block (blue), which subtracts the VCO output
$W_1$ and its sampled and held value $W_{ZOH}$. This generates the continuous-time estimated error signal $E$ which drives the second stage. 
For in-band frequencies, $H_{ZOH}(s) \approx 1$ and $E$ becomes approximately
\begin{equation}
    E\approx-\frac{N_{\phi 1}}{\pi}\cdot Q_1
\end{equation}
Hence the estimated error signal $E$ provides information on the quantization noise of the first stage. Since $E$ is the sum of $N_{\phi 1}$ individual estimated error bits, the frequency gain $K_{2}$ of the second stage must be represented as the gain per individual bit, or $K_{2}= \frac{f_{range,2}}{N_{\phi 1}}$, where $f_{range,2}$ represents the frequency range of the second stage.
The output of the second stage $D_2$ is again a digital signal and can be found analogously to (\ref{eq:y1vco}). The total output signal $D$ can then be calculated by applying the correct NCFs to $D_1$ and $D_2$ and summing the results as shown below. 
\begin{equation}  \left\{
    \begin{array}{ll}
        D(z)=NCF_1(z)\cdot D_1(z)+NCF_2(z) \cdot D_2(z)\\
        NCF_1(z) = 1 \quad \& \quad 
        NCF_2(z) = G \cdot (1-z^{-1})
    \end{array}
\right.
\end{equation}
where $G$ represents the noise cancellation filter gain, which should optimally be $G_{opt}= \frac{f_s}{2 N_{\phi 2} K_2}$. It can then be shown that
\begin{equation} 
    D(z) \approx  \frac{2N_{\phi1}K_1}{f_s} \cdot [V_{in}(s)]^* +G \cdot \frac{N_{\phi 2}}{\pi}\cdot  Q_{2}(z)\cdot(1-z^{-1})^2
    \label{eq:second_order}
\end{equation} 
Equation (\ref{eq:second_order}) shows the expected second-order noise shaping, i.e.~the first-order shaped noise was cancelled. 

A theoretical estimate of the  performance can then be calculated as for a $\Sigma \Delta$-modulator \cite{circ_design} similar to what was done by \cite{vco_kim_jang}. Due to the use of a XOR-based sQSD, both rising and falling VCO edges are counted, allowing $Q_2$ to be modeled as white noise identically distributed between $[0; \frac{\pi}{N_{\phi 2}}[$ \cite{MaghamiJSSC2020}. We first find the signal power as
\begin{equation}
    P_{sig}=\frac{(N_{\phi 1}f_{range,1})^2}{2f_s^2}
\end{equation}
where $f_{range,1} = 2 A K_1 $ represents the frequency range of the first stage, with $A$ the amplitude of the input signal. After integrating over the output-referred noise power spectral density, the in-band noise power can be calculated as
\begin{equation}
    P_Q= \frac{G^2}{12}  \frac{\pi^4}{5 \cdot OSR^5} 
\end{equation}
Finally, the SQNR is found as
\begin{equation}
\begin{split}
    SQNR &= 6.02 \cdot \mathrm{log}_2 \Big ( \frac{
    N_{\phi2} f_{range,1} f_{range,2}}{f_s^2} \Big )\\
    &+ 50 \cdot \mathrm{log}_{10} (OSR) +0.9052
\end{split}
\label{eq:theoretical}
\end{equation}
Though this expression seems to suggest that $N_{\phi 1}$ does not influence the performance, this is only true if $N_{\phi 1}$ is high enough to make the in-band PFM spurs negligible. Additionally, the assumption that the quantization noise $Q_2$ is distributed as white noise is not entirely accurate \cite{gutierrez2017pulse}.

\begin{figure}
    \centering    
    \includegraphics[width=\columnwidth]{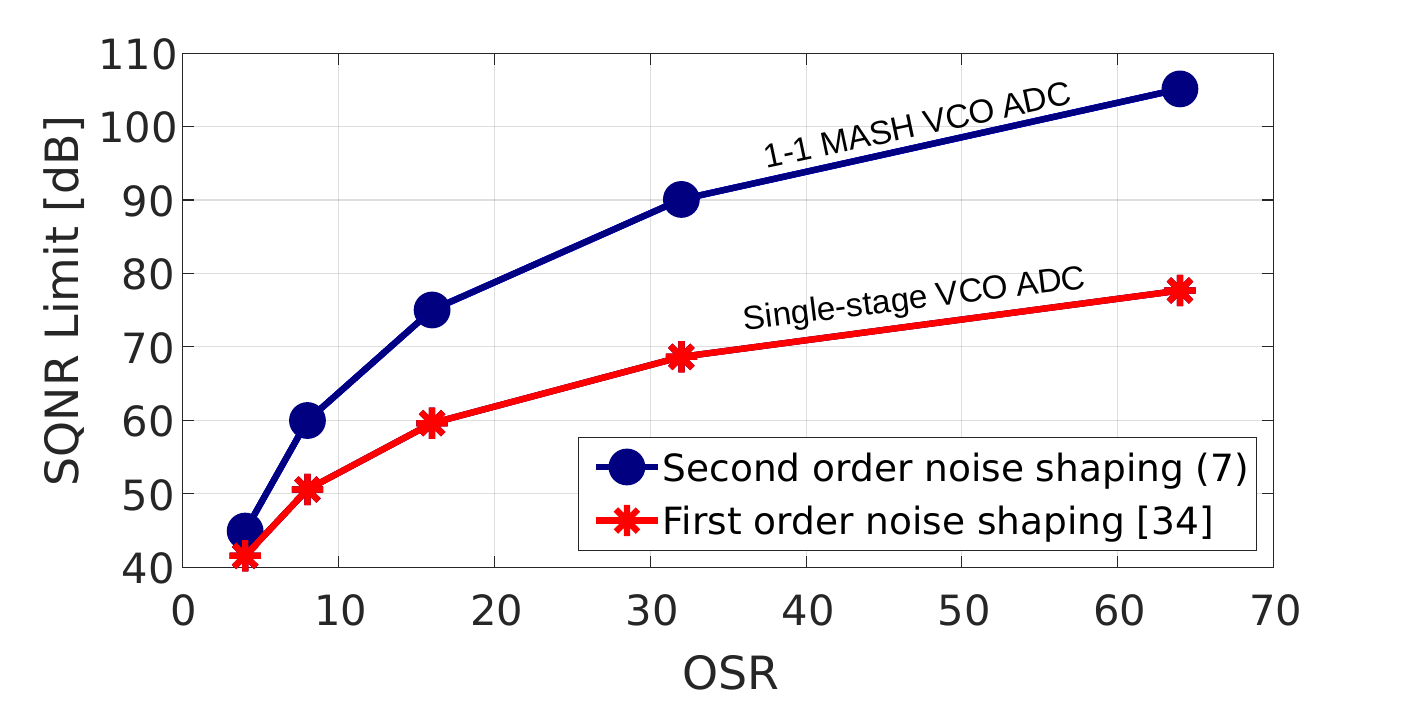}
    \caption{Theoretical SQNR limit in function of the OSR for a single-ended 1-1 MASH VCO ADC with second-order noise shaping or a single-stage first-order VCO ADC using the parameters from Table I.}
    \label{fig:performance_single_vs_mash}
\end{figure}

Building on the results of the previous analysis, we can now compare the performance of a MASH VCO ADC to that of a single-stage VCO ADC. Fig.~\ref{fig:performance_single_vs_mash} presents a comparison between the theoretical SQNR of the MASH architecture as derived in (\ref{eq:theoretical}), and the SQNR of a single-stage VCO ADC, as calculated in \cite{vco_kim_jang} and adapted for a QSD that counts both rising and falling edges. The parameters used for this are summarized in Table I. Realistic values for the frequency range $f_{range}$ and free-running frequency $f_0$ were used, which were based on our final sizing. For the single-stage case, the parameters from the first stage of the MASH were used. Note that this does not include thermal noise or circuit non-idealities, whose influence will be discussed further in this text.

As expected, the benefits of second-order noise shaping increase for higher OSRs, yet remain significant at lower OSRs. This highlights the potential of the MASH-based approach to deliver high performance at increased bandwidths.
\subsection{Comparison With Prior Art 1-1 MASH VCO ADCs}
\begin{figure}[t]
    \center
    \includegraphics[width=\columnwidth]{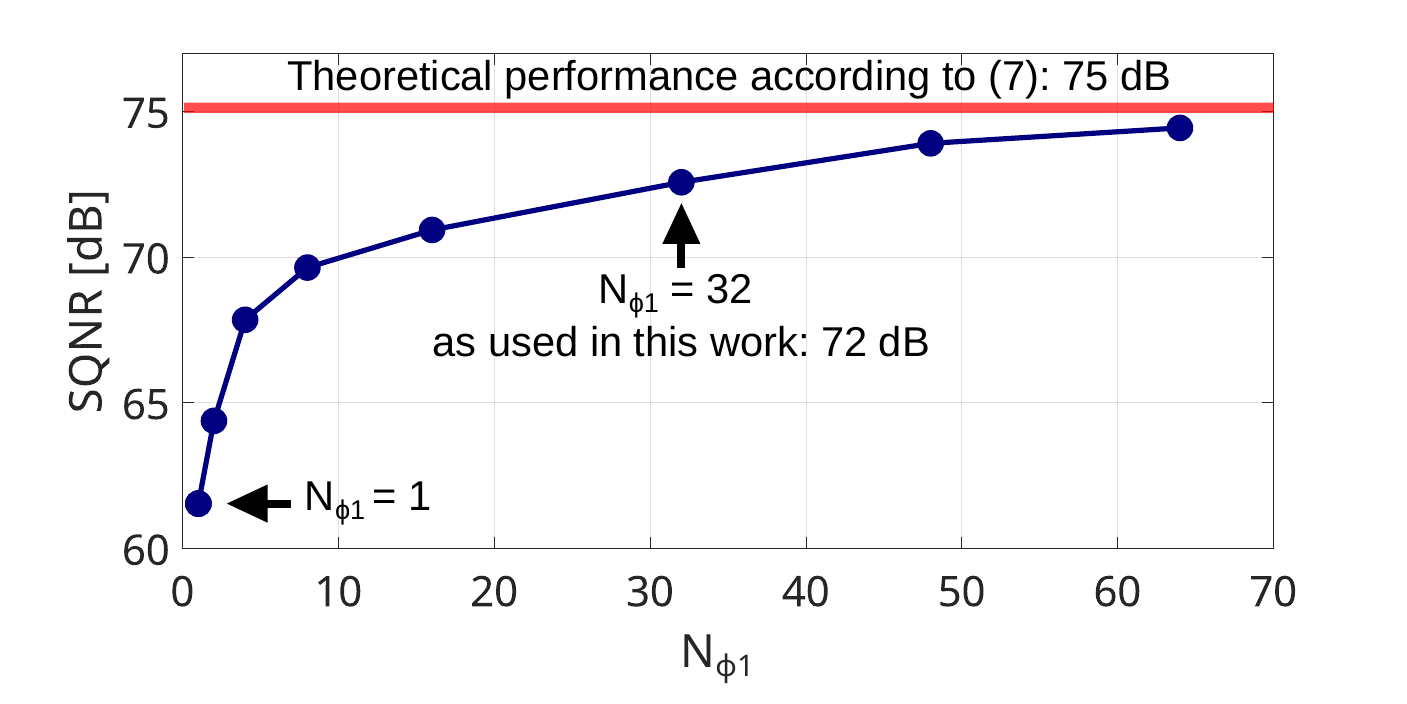}
    \caption{Simulated SQNR of the single-ended 1-1 MASH VCO ADC for different values of $N_{\phi 1}$. The case of $N_{\phi 1}= 1$ corresponds to the architectures of \cite{hernandez_mash, yu_vco}.}\label{fig:sys_snr}
\end{figure} 
\begin{table}%1.21GHz =0.75/0.9*1.45 GHz
\center
    \caption{{Parameters of the 1-1 MASH VCO Architecture.}}\label{table:param}
    \begin{tabular}{|c|c|c|c|}
    \hline
    $f_s$ & 3.5 GHz & OSR & 16 \\ \hline
    $f_{in}$& 31.25 MHz  & $\mathrm{V_{in}}$ & 750 $\mathrm{mV_{pp}}$ \\ \hline
    $N_{\phi 1}$ & 32 & $N_{\phi 2}$ & 32 \\ \hline
    $f_{01}$ & 1.0 GHz& $f_{02}$ & 0.9 GHz  \\ \hline
    $f_{range,1}$ & 1.21 GHz & $f_{range,2}$ & 1.57 GHz \\ \hline
    \end{tabular}
    \end{table}

Previously published work on 1-1 MASH VCO ADCs falls in two categories. In the first category, shortly mentioned above, the coupling is simplified by performing error estimation on a single VCO phase \cite{hernandez_mash, yu_vco}. However, to obtain higher order noise shaping, this limits these MASH ADCs to using only 1-phase readout ($N_{\phi1}=1$) in the first stage, as the quantization error of any additional phases would otherwise not be cancelled. Using the proposed multi-bit error estimation scheme, multi-phase readout in the first stage is possible, leading to improved performance for high bandwidths due to the reduced influence of the in-band PFM spurs.

To determine the advantages of our architecture, we simulated the 1-1 MASH architecture on a system-level using ideal Verilog-A components for a varying $N_{\phi 1}$ using a constant $N_{\phi 2} = 32$. For these simulations, we again use parameters listed in Table \ref{table:param}, with the exception of a variable $N_{\phi1}$ ranging from 1 tot 32. The results of these simulations are shown in Fig.~\ref{fig:sys_snr}.

Assuming a 2-stage design, the case of $N_{\phi 1} = 1$ is equivalent to the systems using single-bit error estimation of \cite{hernandez_mash, yu_vco}. Due to the high bandwidth, the performance for this case is clearly severely limited by the presence of the low-frequency PFM spurs in the spectrum as explained above. As we move towards architectures with multi-bit error estimation, the influence of the PFM spurs decreases and the performance asymptotically approaches the SQNR expected from (\ref{eq:theoretical}).

It could be argued that another readout structure could be more suitable for the $N_{\phi_1}=1$ architectures, as due to the QSDs used in our work,  the maximal VCO frequency is limited to $f_s/2$ \cite{perrott2008}. However, to match our performance, this would imply designing a VCO with a free-running frequency $f_0$ of approximately 32 GHz, which renders the subsequent high-speed counter design far from trivial \cite{baert2020}. The multi-phase approach for the first VCO therefore significantly reduces speed requirements {and facilitates 1-1 MASH VCO ADC conversion at high bandwidths.}

The second category is the architecture implemented in \cite{MaghamiJSSC2020}. Here, multi-phase readout of the first stage is possible by using an approximate 1-bit estimated error signal which takes into account all phases of the first stage. The error is approximated as the delay between the rising edge of the clock and the first subsequent edge of any VCO phase. Due to the closed-loop operation of the first stage, this approximation was shown to be very accurate. 

However, delay mismatches between the error estimation circuits {in detecting the edges} of the different VCO phases will lead to first-order shaped noise leaking into the output spectrum \cite{MaghamiJSSC2020}. First explorations at higher bandwidths revealed these delays become progressively more difficult to match, especially considering the relative complexity of the used error estimation circuits. Additionally, since the pulse lengths of $E$ can be very short in this architecture, a process similar to dead-zone elimination in PLLs is implemented. The error pulses are extended by XOR'ing each sampled phase with the third next phase. It is questionable if this still works at high bandwidths, as this method seems to rely on the assumption that the VCO frequency of the first stage is constant. As the OSR decreases, this frequency will increasingly vary \cite{MaghamiJSSC2020} and the pulse extension will no longer be constant. Due to these reasons, we believe this architecture is not ideal for high-bandwidth operation.
 
\section{Driving the Second Stage\label{sect:cross}}
\subsection{Impact of Nonlinearity in the Second Stage}
\begin{figure}
    \captionsetup{aboveskip=0pt}
    \captionsetup{belowskip=0pt}
    \centering
    \vspace{-0.5em}
    \includegraphics[width=1\columnwidth]{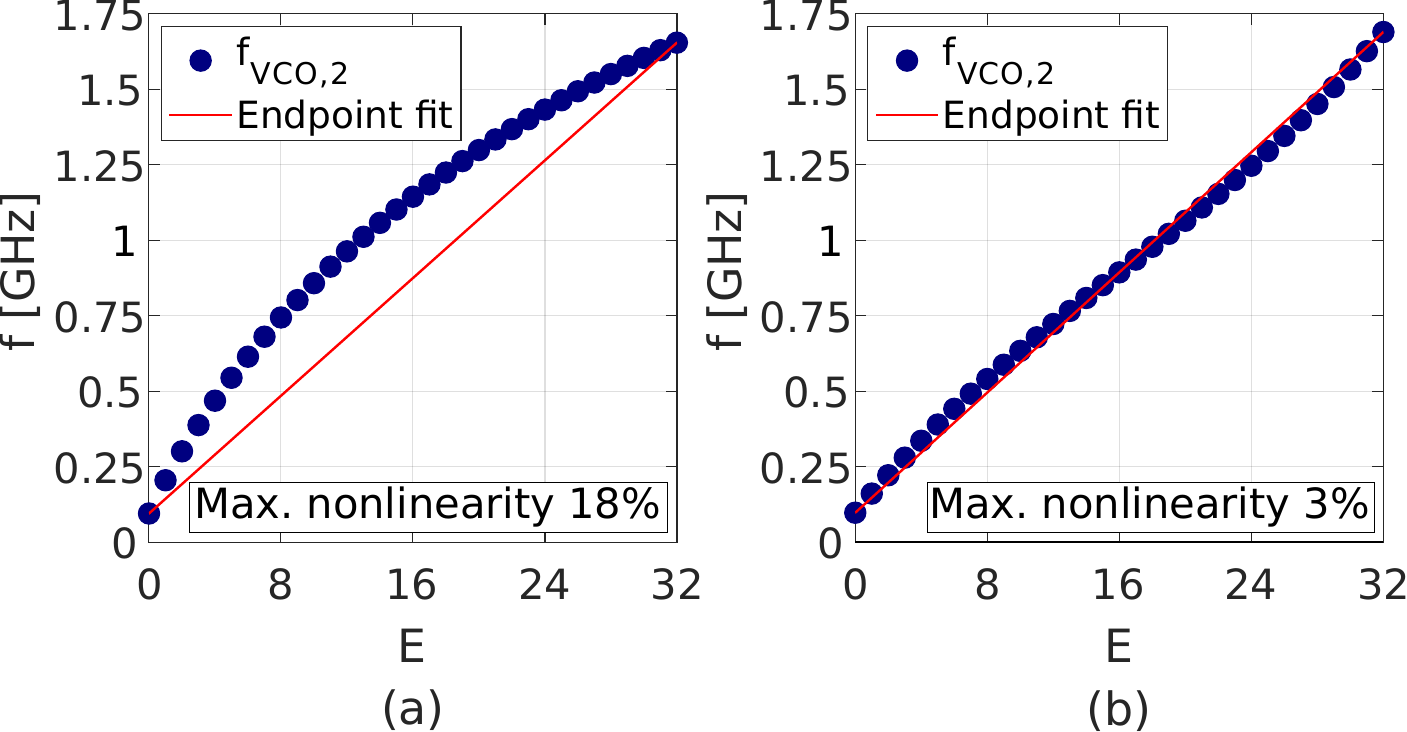}
\caption{Frequency characteristic of the second stage VCO obtained after post-layout simulation without cross-coupling (a) and equivalent with cross-coupling (b). Note that $E$ can only take on integer values between 0 and 32. }\label{fig:sys_freq_dco}
\end{figure}

In Fig.~\ref{fig:sys_freq_dco}(a) the simulated post-layout frequency characteristic of our {{ring oscillator}} in the second stage is shown (blue). The frequency characteristic visibly exhibits mostly second-order nonlinearity.
If a nonlinear ring oscillator is used in the second stage, harmonic distortion of $E$ will be introduced in the system. {From (2), $E$ contains the broadband quantization noise $Q_1$ of the first stage. Distortion of $E$ will therefore add broadband noise which will not cancel out with the quantization noise of the first stage. This effect can be modeled as an added noise source $N_{NL}$ at the input of the second stage as shown in Fig.~\ref{fig:sys_mash} (red).} Equation (\ref{eq:second_order}) now becomes 
\begin{equation}
    \label{eq:NL}
    \begin{split}
        D(z) \approx  &\frac{2N_{\phi1}K_1}{f_s}[V_{in}(s)]^* +{\red G \cdot \frac{2N_{\phi2}K_2}{f_s} N_{NL}(1-z^{-1})}\\ &+G \cdot \frac{N_{\phi 2}}{\pi}\cdot  Q_{2}(z)		\cdot(1-z^{-1})^2 
    \end{split}
\end{equation}
It is clear that a first-order noise shaped component was introduced in the output signal, whose effect on the performance needs to be investigated. 

In order to quantify this, we performed simulations using ideal Verilog-A components with the same parameters as in the previous section. However, the ideal ring oscillator in the second stage was now replaced with its post-layout extraction and driven with ideal current sources, hence realistically modeling the nonlinearity of the second stage.
The resulting SQNR is limited to 69 dB. While this may be a marginally acceptable result for the high-bandwidth design we present here, it leaves very little margin for additional noise sources. In light of this, we explored an architectural improvement, which is the subject of next section. While for our design the improvement mostly serves to create more margin, the new technique is even more important for 1-1 MASH designs that aim for a higher SQNR or in cases where the second stage exhibits more nonlinearity.

\subsection{Architecture With Cross-coupled Estimated Error Signals}
\begin{figure*}
    \center
    \includegraphics[width=\textwidth]{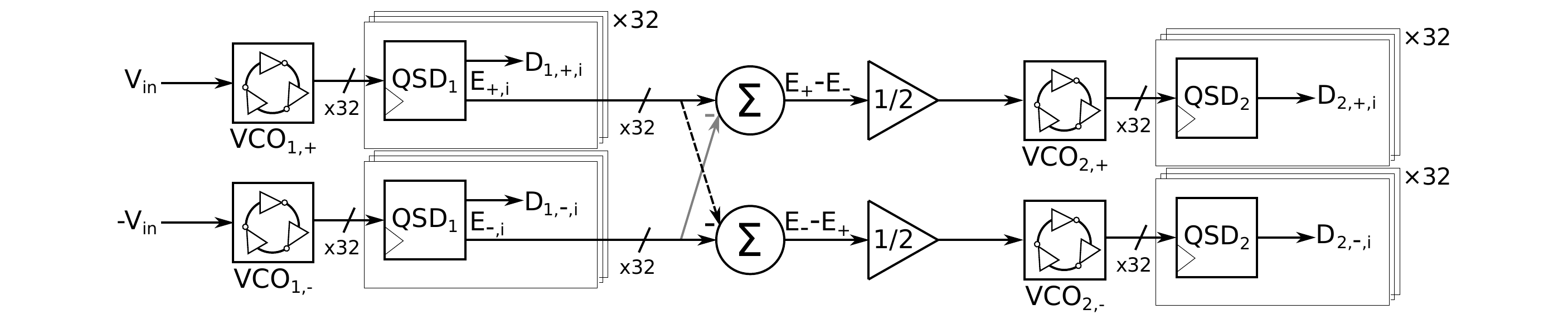}
    \caption{Final improved MASH VCO architecture with cross-coupling.}
    \label{fig:system_good}
\end{figure*}
\begin{figure*}
    \center
    \includegraphics[width=\textwidth]{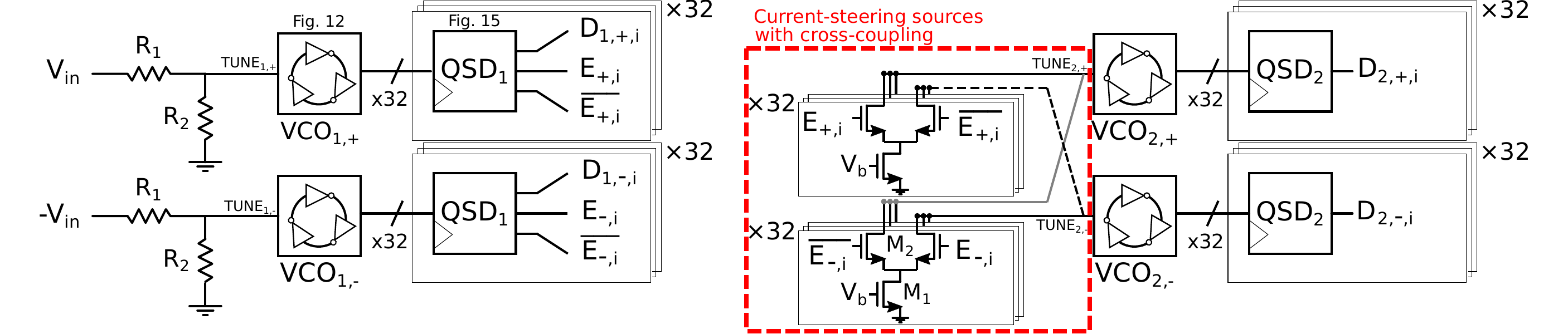}
    \begin{minipage}{0.45\textwidth}
    \center
    \vspace{1\he}
    \begin{tabular}{|c|c|}
    \hline
    $R_1$ & $R_2$\\ \hline
    $150\,\Omega$ & $200\,\Omega$\\
    \hline
    \end{tabular}
    \end{minipage}
    \begin{minipage}{0.45\textwidth}
    \center
    \vspace{1\he}
    \begin{tabular}{|c|c|c|c|c|}
    \hline
    & $W_\mathrm{finger}$ & $\mathrm{\#}$ & $L$ & flavor\\ \hline
    $\mathrm{M}_{1}$ & $300\,\mathrm{nm}$& $18$ & $60\,\mathrm{nm}$ & normal \\ \hline
    $\mathrm{M}_2$ & $200\,\mathrm{nm}$& $2$ & $30\,\mathrm{nm}$ & uhvt \\ \hline
    \end{tabular}
    \end{minipage}
    \caption{Circuit-level view of the proposed implementation with sizing of indicated components.}
    \label{fig:circuit}
\end{figure*}
{VCO ADCs often use a pseudo-differential architecture to reduce even-order distortion. We apply this principle to both stages of the MASH VCO ADC. In a first step, we create a pseudo-differential first stage. This can simply be accomplished by adding a negative channel as shown in Fig.~\ref{fig:system_good} (left). The output $D_1$ of the first stage is then calculated as $D_1 = D_{1,+}-D_{1,-}$. As a consequence, the distortion of the global input signal $V_{in}$ is reduced. 

In contrast with previously published 1-1 MASH VCO architectures \cite{MaghamiJSSC2020, hernandez_mash, yu_vco}, we also use a pseudo-differential setup for the second stage. This will reduce distortion of the broad-band noise contained in $E$ and will therefore mitigate the effect analyzed above. To obtain a pseudo-differential second stage, we first add a negative channel and then propose to cross-couple the estimated error signals of the two channels of the first stage as shown in Fig.~\ref{fig:system_good} (right).
Now the positive and negative channel are driven by the opposite signals $E=E_+-E_-$ and $-E=E_--E_+$ respectively. The rationale for this choice of signals is that due to the pseudo-differential operation of the first stage, the total quantization noise of the first stage now takes the form $Q_{1,+}-Q_{1,-}$, which must be present in the output of the second stage to obtain second-order noise shaping. A division by 2 is introduced to keep the input values of each channel of the second stage in the same range. By calculating $D_2$ as $D_2 = D_{2,+}-D_{2,-}$, we have created a pseudo-differential second stage. 

Simulations were performed for the cross-coupled architecture using the same parameters as above. The SQNR without nonlinearity is now increased to 75 dB due to the pseudo-differential operation. When performing simulations including nonlinearity, an SQNR of 74 dB is found, which only represents a reduction of 1 dB compared to the ideal case. The cross-coupling has therefore led to a reduced impact of nonlinearity in the second stage. Notice that there is still a small reduction in performance, as the pseudo-differential operation does not suppress uneven-order distortion and some quantization noise leakage therefore still occurs.

The effect of cross-coupling is also visible in the equivalent frequency characteristic of the second stage after cross-coupling, which is shown in Fig.~\ref{fig:sys_freq_dco}(b). This clearly demonstrates the linearized characteristic and the suppression of the dominant second-order nonlinearity compared to Fig.~\ref{fig:sys_freq_dco}(a). The maximum nonlinearity was reduced from $18\%$ to $3\%$.

An additional advantage of the cross-coupling is that the entire architecture now follows the guidelines outlined in \cite{borgmans2022mismatch}, which recommends the use of a pseudo-differential setup for all ring oscillators to improve the power-supply rejection ratio (PSRR) and  common-mode rejection ratio (CMRR).
A final system-level advantage of the cross-coupling will be explained in section \ref{subsect:pulse}, where it will be shown that using a pseudo-differential second stage leads to an increased robustness against pulse-width errors in the error estimation circuit. 
}
\subsection{Practical Implementation}
As explained above, the second stage is operated pseudo-differentially by cross-coupling the estimated error signals $E_+$ and $E_-$ from the positive and negative channel of the first stage. In our practical implementation, the combination of the $E_i$ bits and cross-coupling are performed in a natural way by driving the second stage by an array of differential current-steering current sources operated by the bits $E_{i}$ and $\overline{E_{i}}$ as shown on the right of Fig.~\ref{fig:circuit}. These current-steering sources are common in high-speed DACs and offer a highly improved dynamic behavior compared to single-ended alternatives \cite{razavi_2018}. The use of a pseudo-differential second stage allows for the efficient use of these current-steering sources, as all current used for the combination of the $E_i$ bits is also used to drive the ring oscillators of the second stage. To operate the current sources, the error estimation is expanded to generate both $E_{i}$ and its complementary signal $\overline{E_{i}}$. The bits $E_{i}$ and $\overline{E_{i}}$ are generated in parallel using symmetrical logic gates to obtain the same timing behavior, without requiring any further logic. The implementation for this is discussed in section \ref{subsect:qsd}.

\subsection{Influence of Pulse Width Errors  \label{subsect:pulse}}
\begin{figure}[t!]
    \centering
    \includegraphics[width=1\columnwidth]{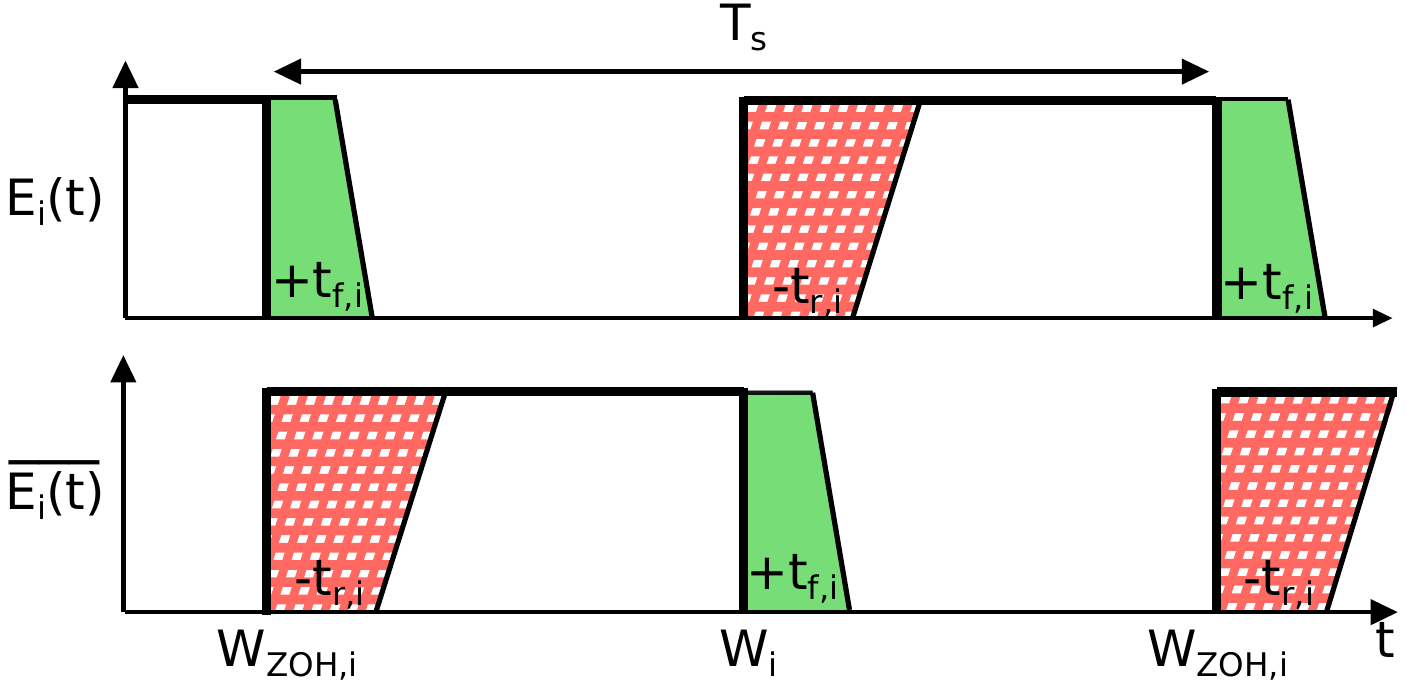}   
    \caption{Conceptual illustration of the influence of pulse-width errors.} \label{fig:cir_pulse}
\end{figure}
{As the information contained in $E_i$ is contained in the pulse widths, an important concern in 1-1 MASH VCO ADCs is the presence of pulse width errors in the error estimation circuit \cite{MaghamiJSSC2020}.} This issue can be understood as follows. Let us consider the signal $E_i$ and its complement $\overline{E_i}$. These are outputs of an error estimation circuit with a VCO phase $W_i$ and its sampled and held value $W_{ZOH,i}$ as inputs. The time-domain waveforms are shown in Fig.~\ref{fig:cir_pulse}. Ideally, $E_i$ would go down at the rising edge of the sampled and held $W_{ZOH,i}$ and go up at the rising edge of $W_i$, as indicated. The inverse is true for $\overline{E_i}$. The ideal pulse width of these signals is shown in bold lines. However, realistically, the rising edges will be delayed and undergo a limitation on their rise time. We indicate this with an effective delay $t_{r,i}$. The delay and rate limitation of the falling edges will be modeled by a similar $t_{f,i}$. The effect of $t_{r,i}$ and $t_{f,i}$ is that a current proportional to $t_{f,i}-t_{r,i}$ will be added to the signal to the second stage during a sampling period. Moreover, every $W_i$ can have a different $t_{f,i}-t_{r,i}$, e.g.~due to path length variations in the layout when connecting to the second stage. Other factors that can contribute to this include proximity effects and systemic variations due to temperature or stress gradients \cite{Pelgrom2017}. In the initial single-ended architecture, only $E_i$ is used as input of the second stage, which leads to an introduction of mismatch noise in the input of the second stage as the VCO cycles through the different $E_i$ bits \cite{MaghamiJSSC2020}. This is especially relevant for high-speed designs, as the relative duration of pulse width variations compared to the clock period increases for higher clock frequencies.

In our implemented architecture with cross-coupling, we expect this effect to be mitigated. Due to a symmetrical layout, path length variations of $E_i$ and $\overline{E_i}$ of a single error estimation circuit will be very similar. Additionally, these signals will be subjected to comparable proximity effects and gradients. As a result, $t_{f,i}-t_{r,i}$ of $E_i$ and $\overline{E_i}$ can be matched. The same current proportional to $t_{f,i}-t_{r,i}$ will be added to both the positive and negative channel of the second stage. These contributions will largely compensate each other due to the pseudo-differential operation of the second stage, thereby reducing the effect of $t_{f,i}-t_{r,i}$ variations. 

To demonstrate the effect of cross-coupling, we again perform simulations using ideal Verilog-A components, comparing the single-ended setup with the proposed design. In order to model the effect discussed above, delays and rise and fall limitations were added to the error estimation circuit. To stress test the robustness of both designs, relatively large variations were introduced with $t_{r,i}-t_{f,i}$ distributed from 0 ps to 75 ps in a gradient-like fashion, to reflect path length variations in the wiring towards the second stage.

For the single ended-configuration, this simulation results in an SQNR of 65 dB, or a reduction of 7 dB compared to the ideal single-ended results of 72 dB using only Verilog-A components obtained above. For our proposed architecture using cross-coupling, an SQNR of 71 dB is obtained, which is a reduction of 4 dB compared to the ideal results using cross-coupling. The effect of $t_{f,i}-t_{r,i}$ variations has therefore been reduced, leading to a significant improvement in SQNR. This increased robustness, even for very the large $t_{f,i}-t_{r,i}$ variations in this stress test, is an interesting advantage of the proposed cross-coupled architecture. 

\section{CMOS Circuits\label{sect:cir}}
A block diagram of the circuit is shown in Fig.~\ref{fig:circuit}. The different building blocks will be described in the next sections. The circuits were developed for a 28nm CMOS technology, which offers transistors in different flavors, including ultra low/high treshold voltage transistors (ulvt/uhvt). Note that in sizing tables, the symbol \# represents the total multiplier for each finger (ie.~the number of fingers per transistor times the multiplier of each transistor).

\subsection{Ring Oscillators} \label{subsect:ring_oscillator}
\begin{figure}
    \center
    \flushleft{(a)}    \vspace{-1.5\he}
    \center
    \includegraphics[width=\columnwidth]{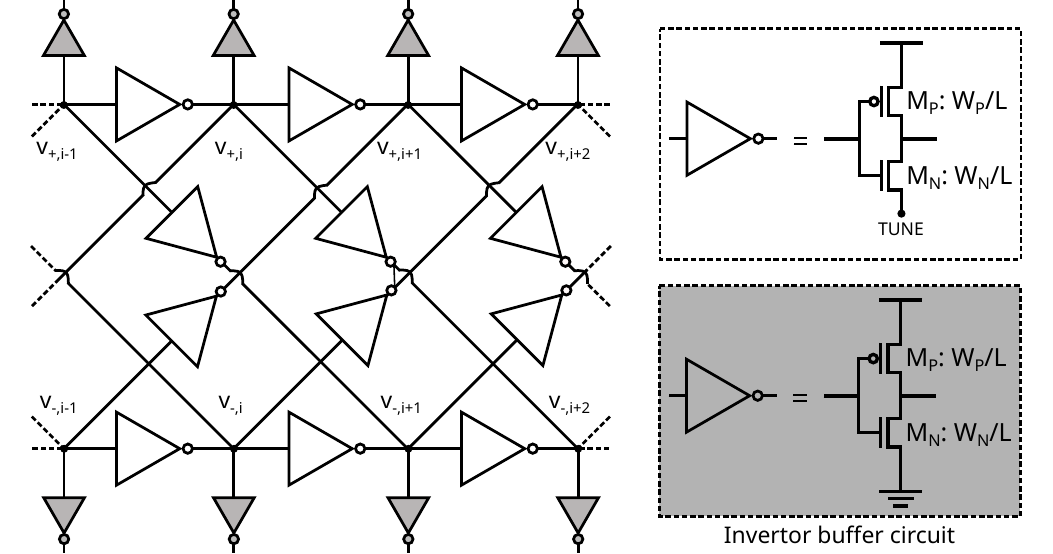}
    \flushleft{(b)}    \vspace{-1.5\he}
    \center
    \begin{tabular}{|c|c|c|c|c|c|}
    \hline
    $W_{N,\mathrm{finger}}$ & $W_{P,\mathrm{finger}}$ & L & $\mathrm{\#_{stage1}}$&$\mathrm{\#_{stage2}}$\\ \hline
    $200\,\mathrm{nm}$ & $400\,\mathrm{nm}$ & $30\,\mathrm{nm}$ &16 &2\\
    \hline
    \end{tabular}
    \flushleft{(c)}    \vspace{-1.5\he}
    \center
    \begin{tabular}{|c|c|c|c|c|}
    \hline
    & $W_\mathrm{finger}$ & $\mathrm{\#}$ & $L$ & flavor\\ \hline
    $\mathrm{M}_N$ & $100\,\mathrm{nm}$& $1$ & $30\,\mathrm{nm}$ & uhvt \\ \hline
    $\mathrm{M}_P$ & $400\,\mathrm{nm}$& $1$ & $30\,\mathrm{nm}$ & ulvt \\ \hline
    \end{tabular}
    \caption{Ring oscillator using feed-forward delay cells (a) along with the sizing of the delay cells in the first \& second stage (b), sizing of the invertor buffers (c).}
    \label{fig:cir_ring}
\end{figure}

The ring oscillator VCOs consist of 32 differential feed-forward delay cells, shown in Fig.~\ref{fig:cir_ring}(a)   \cite{borgmans2019enhanced,baert2020}. Each invertor is identically sized. This topology was selected as it is better suited to high-bandwidth operation than a traditional structure using cross-coupled auxiliary invertors \cite{borgmans2019enhanced}. 

The first stage is directly driven by the overall input voltage $V_{in}$ through the resistive drive circuit of \cite{nguyen2021deep, amir_2016,borgmans2019enhanced}, shown on the left of Fig.~\ref{fig:circuit}, which will be combined with an off-chip calibration procedure to achieve our desired linearity. This procedure is further described in section \ref{subsect:cal}.

As was explained above, the cross-coupling and summation are performed by driving the second stage by an array of differential switched current sources operated by the bits $E_{i}$ and $\overline{E_{i}}$. The switched-current sources are shown in Fig.~\ref{fig:circuit} (red) along with their sizing.

The ring oscillator sizing is based on thermal noise considerations using equations developed in \cite{borgmans2021analog,borgmans2022noise}. These are 
\begin{equation*}
    \begin{split}
        S_{w,\mathrm{in,VCO}} &\approx 4 kT R_\mathrm{ring} \left(1+\frac{R_1}{R_2}\right)^2 \frac{\gamma_n+\gamma_p}{2} \\
        S_{w,\mathrm{in},R} &= 4 kT R_1 \left(1+\frac{R_1}{R_2}\right)
    \end{split}
\end{equation*} 
Where $\gamma_N$ and $\gamma_P$ are the NMOS and PMOS channel noise factors respectively. $R_\mathrm{ring}= \frac{\partial V_\mathrm{ring}}{\partial I_\mathrm{ring}}$ can be extracted using the I(V)-characteristic of a simple delay cell where the transistors consist of 1 finger and evaluating it at the intersection with the load line of the resistive driver as explained in \cite{borgmans2022noise}. 

The ratio of the resistors $R_1/R_2$ in the drive circuit is a compromise between the linearity and the frequency tuning range, which determines the SQNR. In this design, the SQNR is prioritized, while we count on the end-of-chain calibration to reduce the distortion components. However, because calibration becomes harder as nonlinearity increases, a target uncalibrated nonlinearity of around 1\% has been put forward, which we believe is sufficient for good calibration. This leads to the sizing of the $R_1/R_2$ ratio equal to 0.75. $R_{ring}$ can be decreased by increasing the amount of fingers of the transistors in the delay cell.

Since the thermal noise introduced by the second stage is heavily suppressed by the VCO gain $K_{1}$ of the first stage and shaped  when referred to the input, the second ring oscillator can be much smaller. We opt for a width 4 times larger than minimal to avoid bad PSSR \cite{borgmans2022mismatch}.

\begin{figure}
    \centering
    \includegraphics[width=1\columnwidth]{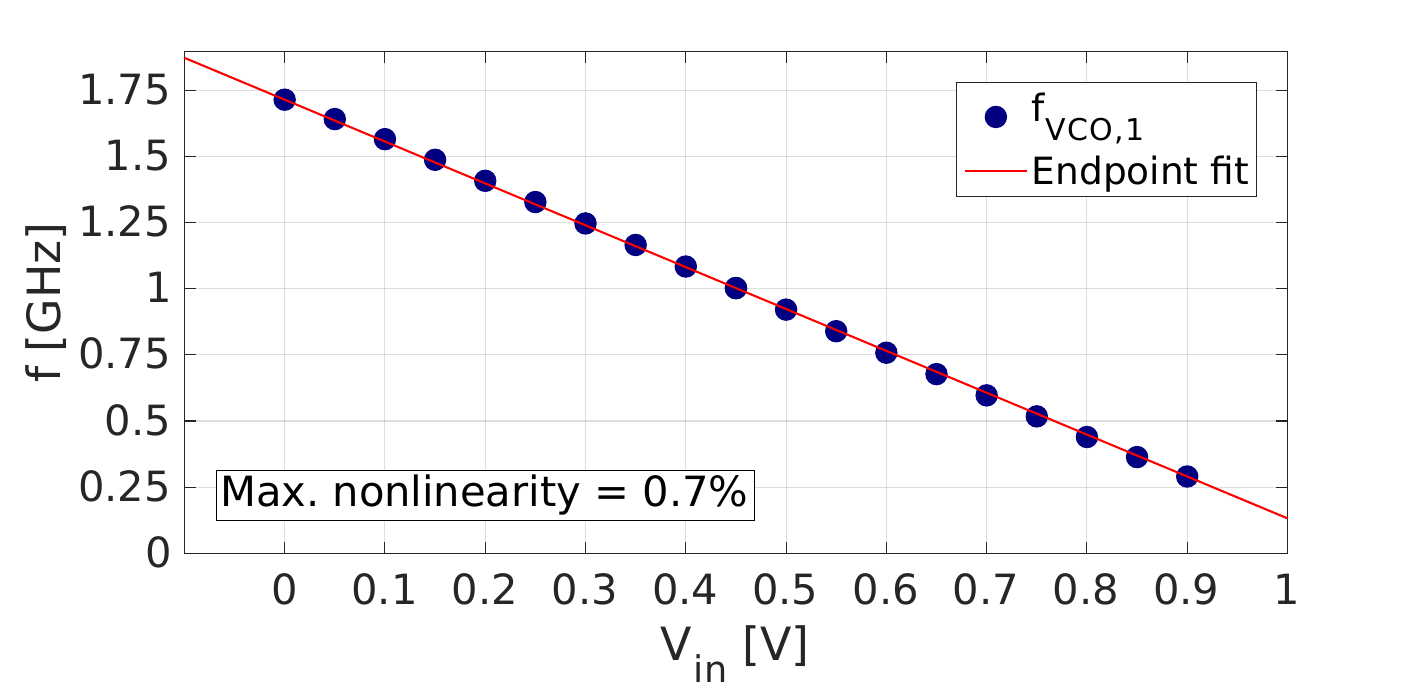}   
    \caption{Frequency characteristic of the first stage VCO obtained after post-layout simulation.} \label{fig:freq_vco}
\end{figure}

\begin{figure}
    \centering
    \includegraphics[width=1\columnwidth]{{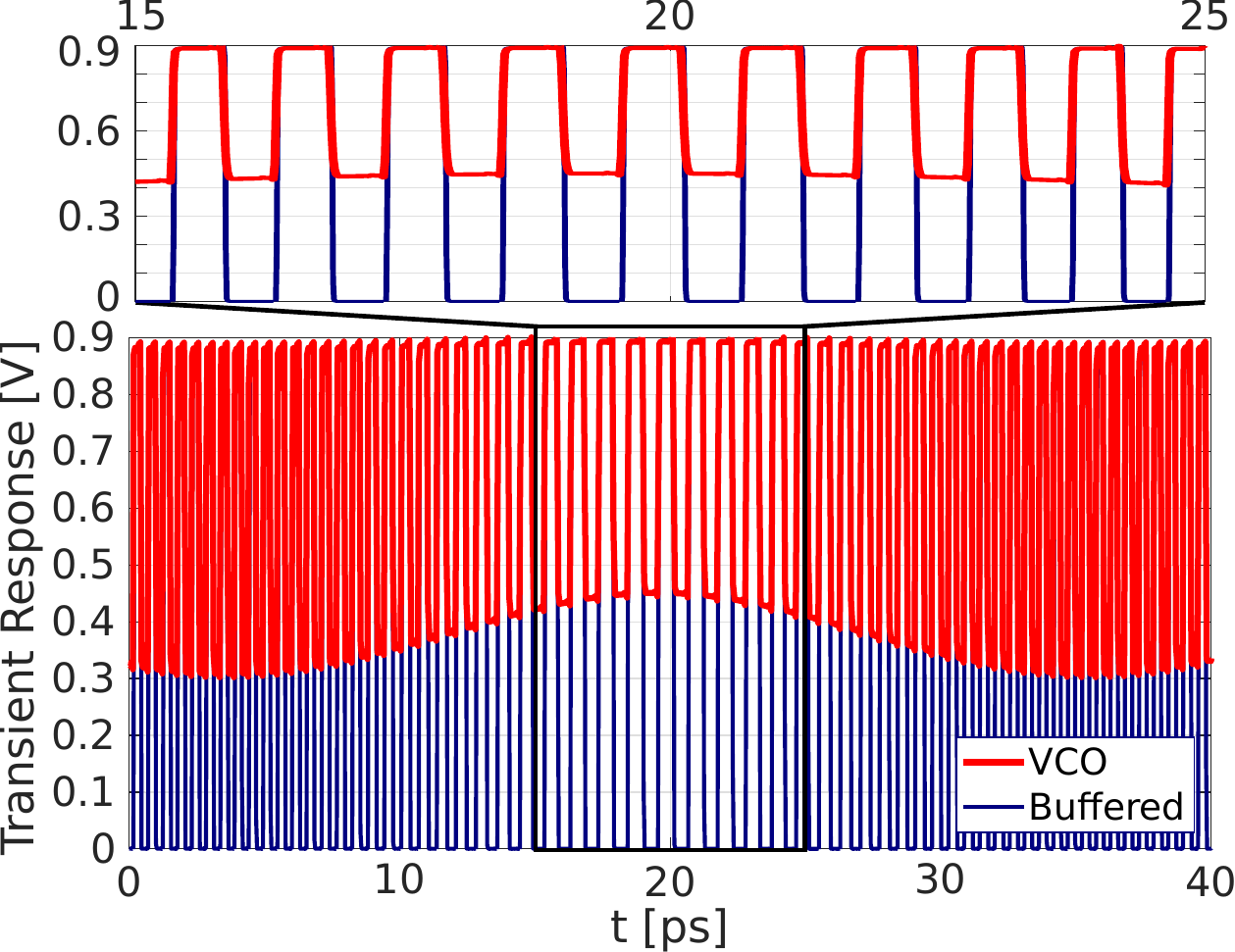}}
    \caption{Post-layout output waveform of a VCO delay cell and the subsequent buffer for a 750 $\mathrm{mV_{pp}}$ 26.5 MHz input.}\label{fig:vco_buff}
\end{figure}

The ring oscillators are sized in order to obtain an SNR due to only thermal noise of 71 dB, leading to the sizings in Fig.~\ref{fig:cir_ring} and frequency characteristics in  Fig.~\ref{fig:freq_vco} and Fig.~\ref{fig:sys_freq_dco} for the ring oscillators in the first and second stage respectively. With the post-layout simulated SQNR also at 71 dB, a total SNR of 68 dB is projected.
Invertor buffers (gray invertors in Fig.~\ref{fig:cir_ring}) are introduced between the VCOs and the QSDs for two reasons. First, they serve to isolate the VCOs from kickback from the sense amplifiers.
Second, the sense amplifier inputs must be rail-to-rail to avoid signal-dependent delays.  The buffers are sized to pull the VCO outputs rail-to-rail while minimizing the capacitive load on the VCOs. Their sizing is summarized in Fig.~\ref{fig:cir_ring}(c). The output of a VCO delay cell is shown in Fig. \ref{fig:vco_buff} before (red) and after the buffers (blue). It can be observed that the buffers successfully pull the VCO output rail-to-rail.
\subsection{QSD With Error Estimation}\label{subsect:qsd}
\begin{figure}
    \center
    \includegraphics[width=1\columnwidth]{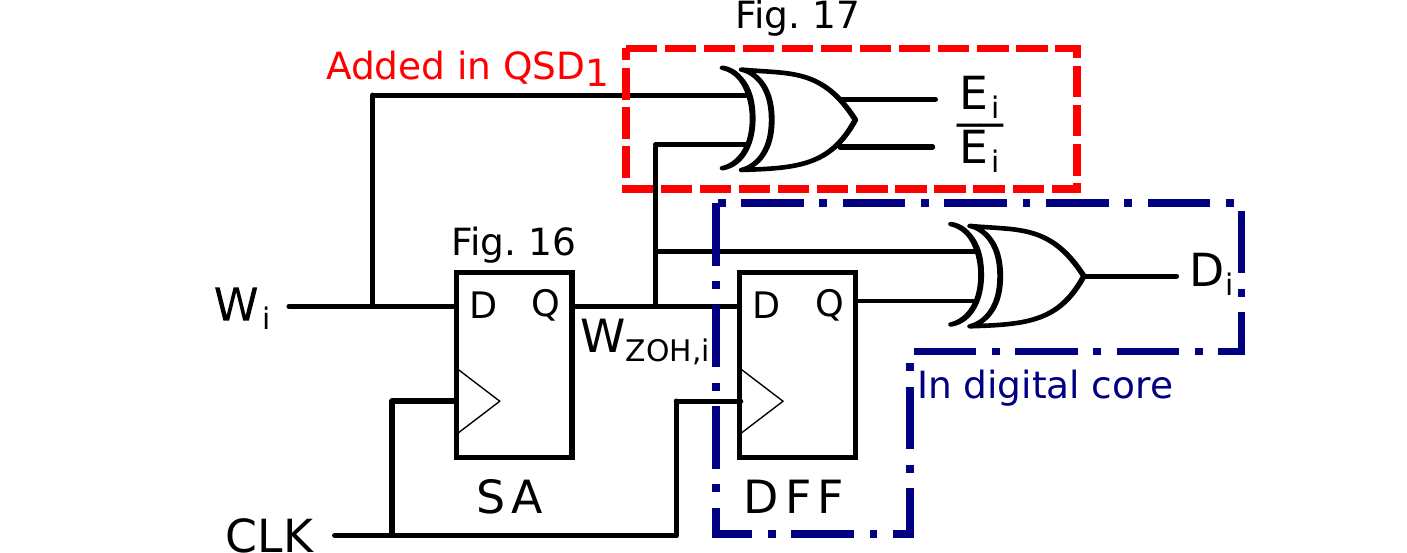}
    \caption{Simplified single-ended representation of the QSD block with error estimation.} \label{fig:cir_qsd}
\end{figure}

A key building block is the QSD following each VCO phase, shown in Fig.~\ref{fig:cir_qsd} \cite{gutierrez2017pulse}. The
first flipflop of the QSD is implemented as a sense amplifier for reasons that will be explained in the next paragraph. The second flipflop and XOR gate implement the differentiation and are embedded in the automatically synthesized digital core.

\begin{figure}
    \center
    \flushleft{(a)}       \vspace{-2.5\he}
    \center
    \includegraphics[width=1\columnwidth]{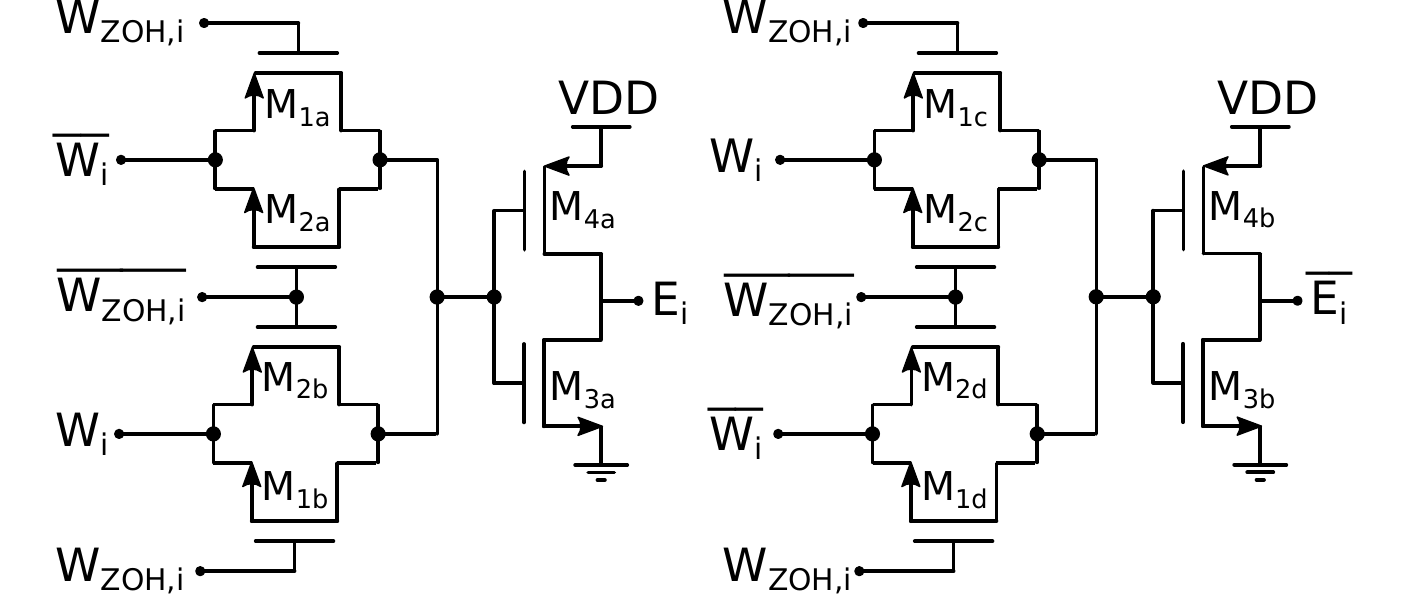}
    \flushleft{(b)}       \vspace{-2.5\he}
    \center
    \begin{tabular}{|c|c|c|c|c|}
    \hline
    & $W_\mathrm{finger}$ & $\mathrm{\#}$ & $L$ & flavor\\ \hline
    $\mathrm{M}_1$-$\mathrm{M}_3$ & $100\,\mathrm{nm}$& $1$ & $30\,\mathrm{nm}$ & ulvt \\ \hline
    $\mathrm{M}_4$ & $300\,\mathrm{nm}$& $2$ & $30\,\mathrm{nm}$ & ulvt \\ \hline
    \end{tabular}
    \caption{Pass-transistor XOR gate  (a) and sizing (b).} \label{fig:cir_xor}
\end{figure}

For the error estimation, a XOR operation is added to the QSD of the first stage, shown in red in Fig.~\ref{fig:cir_qsd}. This implements the substraction shown in Fig.~\ref{fig:system_naive}. The XOR gates are shown in Fig.~\ref{fig:cir_xor} along with their sizing. Fast operation is achieved by using a pass-transistor logic implementation. {Symmetrical circuits are used to generate $E_i$ and $\overline{E_i}$ in parallel.} These 1-bit signals are then directly connected to the cross-coupled switched current sources driving the second stage.
\begin{figure}
    \center
    \flushleft{(a)}    \vspace{-2.5\he}
    \center
    \includegraphics[width=1\columnwidth]{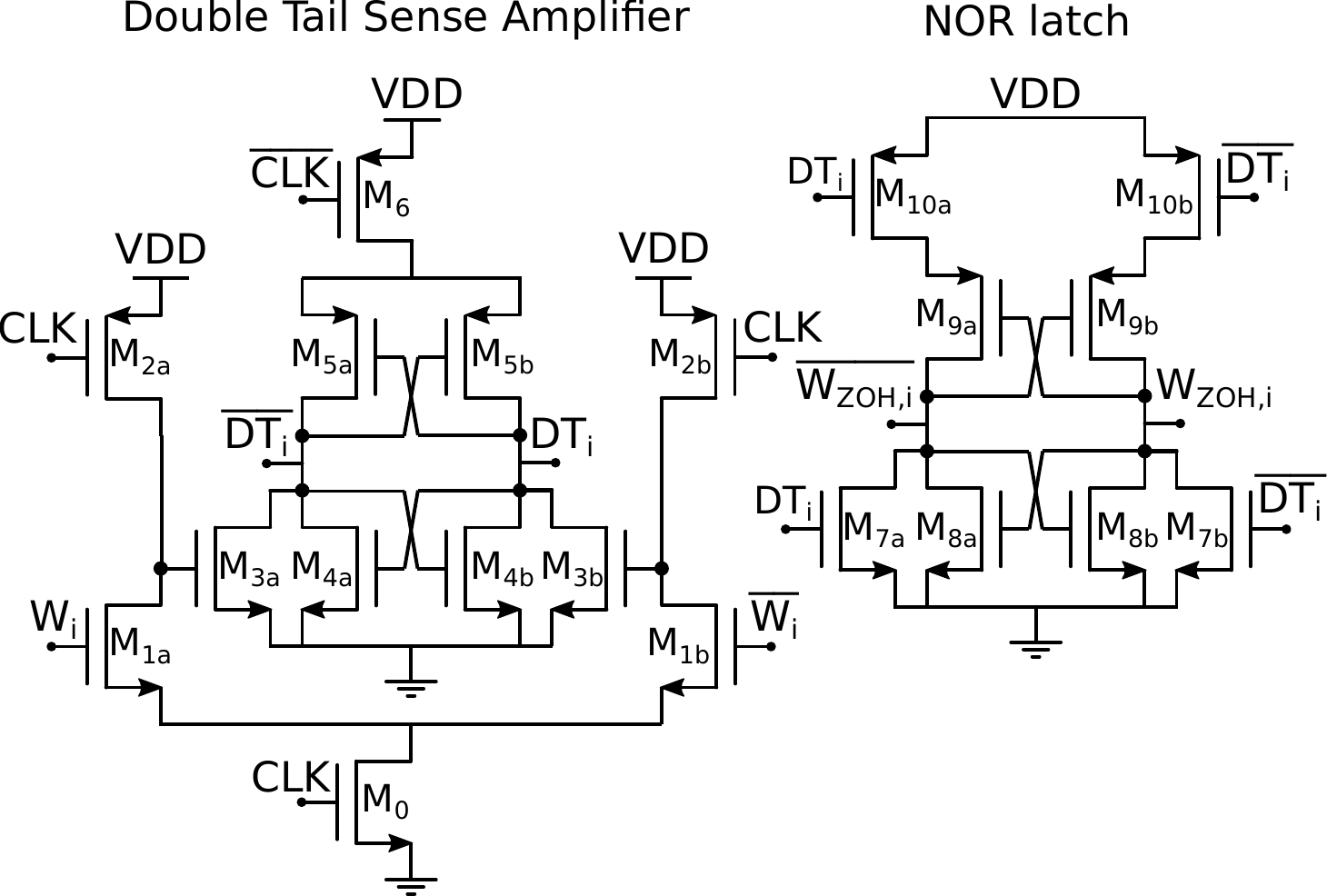}
    \flushleft{(b)}    \vspace{-2.5\he}
    \center
    \begin{tabular}{|c|c|c|c|c|}
    \hline
    & $W_\mathrm{finger}$ & $\mathrm{\#}$ & $L$ & flavor\\ \hline
    $\mathrm{M}_0$ & $200\,\mathrm{nm}$& $2$ & $30\,\mathrm{nm}$ & ulvt \\ \hline
    $\mathrm{M}_1$ & $300\,\mathrm{nm}$& $1$ & $30\,\mathrm{nm}$ & ulvt \\ \hline
    $\mathrm{M}_6$ & $100\,\mathrm{nm}$& $2$ & $30\,\mathrm{nm}$ & ulvt \\ \hline
    $\mathrm{M}_3$, $\mathrm{M}_{10}$ & $200\,\mathrm{nm}$& $1$ & $30\,\mathrm{nm}$ & ulvt \\ \hline
    $\mathrm{M}_2$, $\mathrm{M}_{4-5}$, $\mathrm{M}_{7-9}$ & $100\,\mathrm{nm}$& $1$ & $30\,\mathrm{nm}$ & ulvt \\ \hline
    \end{tabular}
    \caption{Double-tail sense amplifier (left) and latch to maintain output over entire clock period (right) (a) and sizing (b).}\label{fig:cir_dt}% The symbol \# represents the total multiplier for each finger.}  
\end{figure}

For high-speed ADCs in general, metastability is an important concern. In the MASH architecture, this is also highly relevant. As mentioned above, information about $Q_1$ is contained in the pulse lengths of $E_i$. 
There is a finite chance that a transition of a VCO phase $W_i$ happens close to the rising edge of the clock. Consequently, a relatively small signal is applied to the sense amplifier, leading to a metastability event. In this case, the rising edge of $E_i$ will be delayed by the time necessary for the sense amplifier to amplify the input to a logic value. $Q_1$ will not be cancelled completely and first-order shaped noise will leak into the spectrum. The need for a sufficiently fast sense amplifier was confirmed through system-level simulations. 

After a comparison of multiple sense amplifier topologies, the double tail sense amplifier was selected. This is a regenerative sense amplifier offering a very small regeneration time and consequently, an extremely small metastability window \cite{xu2019analysis}.
Since the double-tail sense amplifier has a reset operation when $CLK$ goes low, its outputs are invalid during half a clock period. The sense amplifier is therefore followed by a set-reset latch to maintain its output during the entire clock cycle \cite{xu2019analysis}, thereby implementing the desired sample and hold behavior. A cross-coupled NOR latch is used. The sense amplifier and latch are displayed in Fig.~\ref{fig:cir_dt} along with their corresponding sizing. Through simulations using a post-layout extracted sense amplifier, it was assessed that the sense amplifier speed was sufficient to avoid performance loss due to metastability.

\subsection{Digital Core}
\begin{figure}
    \center
    \includegraphics[width=1\columnwidth]{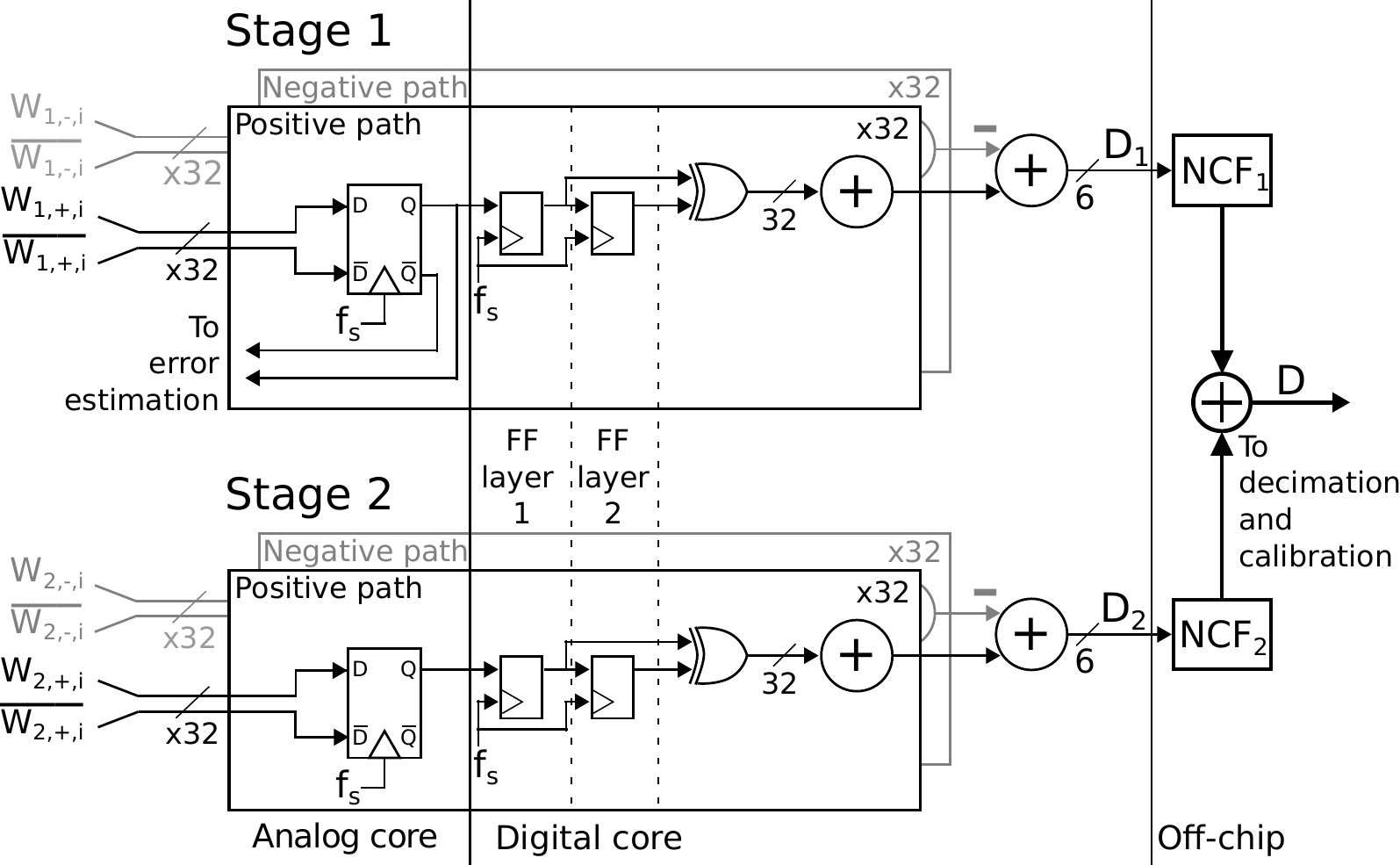}
    \caption{Signal flow diagram in the digital core.}
    \label{fig:analog_digital}
\end{figure}

The digital nature of VCO ADC output signals allows to fully exploit the use of automated tools for their subsequent processing. In this work, the use of these tools was also explored. For this, the synthesis was generated  from a behavioral Verilog description through Cadence's Genus Synthesis Solution. For the automated place and route the Innovus Implementation System was used. This tool can also automatically insert signal and clock tree buffering. Configuration scripts were written for both. Unfortunately, due to lack of experience with these complex tools, the resulting digital circuit, while fully functional, is less efficient than we hoped for. For example, a layer of power-consuming and unnecessary registers was inserted (layer 1) as visible in Fig.~\ref{fig:analog_digital}, which shows the signal flow through the digital core. We only discovered this after fabrication where we measured a higher-than-expected power consumption in the digital circuit.

Note that the digital core does not include the NCFs, which are implemented off-chip as is often done for experimental MASH ADCs \cite{MaghamiJSSC2020, liu2022158, meng_2023}. However, we found it important to include the estimated power consumption and area in case of an on-chip implementation. After place and route with the aforementioned automated tools, an added power of 1.5 mW and area of 0.0006 $\mathrm{mm^2}$ were found.
\subsection{Inter-stage Gain Mismatch}
Multiple sources of non-idealities in the circuit were already discussed above. For example, the influence of nonlinearity in the second stage, pulse width errors and metastability in the sense amplifiers were addressed. A final factor that must be considered is inter-stage gain mismatch, which can also have an impact on the performance of the MASH.

To quantify the effect of inter-stage gain mismatch, we first performed post-layout simulations to evaluate the range of possible values for the optimal noise cancellation filter gain $G_{opt}$ over process and temperature corners. Since $G_{opt}=\frac{f_s}{2 N_{\phi 2} K_2}$, as explained in section \ref{subsect:analysis}, $G_{opt}$ can be calculated in function of the frequency gain $K_2$ of the second stage. The $G_{opt}$ values resulting from these simulations are summarized in Table \ref{tab:G_range}. This table also shows the deviation with respect to the $G_{opt}$ value obtained under nominal conditions (27$^{\circ}$C TT). The corners with the largest $G_{opt}$ deviations are indicated in bold and are found for 0$^{\circ}$C SS (13 \%) and 70$^{\circ}$C FF (-13 \%). As a consequence, by using the value of $G_{opt}$ obtained under nominal simulation conditions, we do not expect larger $G_{opt}$ deviations than $\pm$13\% over temperature and process corners. 
 
\begin{figure}
    \centering
    \includegraphics[width=1\columnwidth]{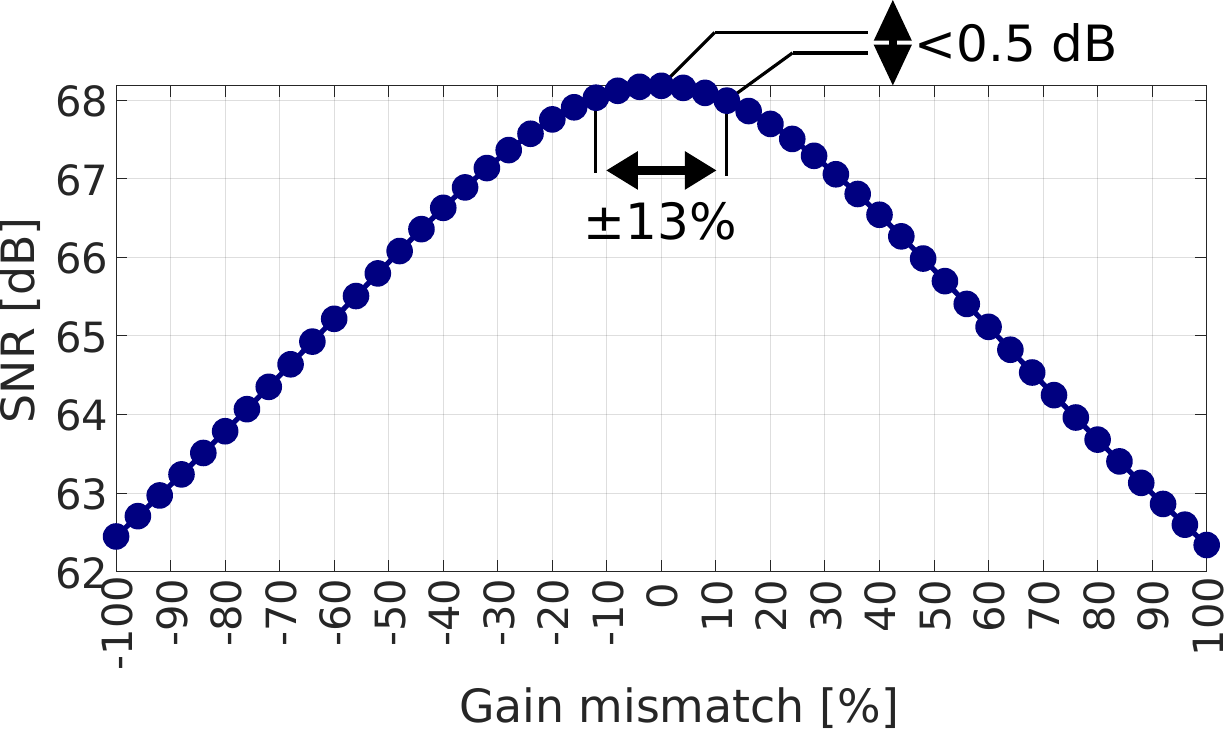}   
    \caption{{Influence of variations of the noise cancellation gain $G$ compared to the optimal value $G_{opt}$ on the post-layout simulated SNR including thermal noise.}} \label{fig:cir_gain}
\end{figure}

Afterwards, post-layout simulations including thermal noise were performed to evaluate the effect of mismatch of $G$ compared to its optimum. The results are shown in Fig.~\ref{fig:cir_gain}. Note that the case of -100\% gain mismatch corresponds to an inter-stage gain of 0, which therefore represents the performance of a single-stage VCO ADC consisting of only the first stage. From Fig.~\ref{fig:cir_gain}, we find that the architecture can tolerate variations of $G$ of the aforementioned $\pm 13\%$ around its optimum with less than a 0.5 dB loss in SNR. This relatively flat behavior around the optimum can be understood by considering that even with a small amount of first stage quantisation noise leakage due to non-ideal gain matching, the total noise at this point will still be dominated by other contributions, most notably the thermal noise. Additionally, eventual quantization noise leakage will still be first-order noise shaped. Only when the gain mismatch exceeds a certain level, we notice a rapidly decreasing SNR. Qualitatively, these observations are also consistent with the measurement results of the 1-1 MASH VCO ADC reported in \cite{MaghamiJSSC2020}. 

By using the value of $G_{opt}$ obtained under nominal post-layout simulation conditions, we therefore do not expect deviations larger than $\pm$13\% over temperature and process corners. This corresponds to an expected reduction in SNR due to NTF gain mismatch of less than 0.5 dB. We found this acceptable and did not use a separate $G$ calibration, but used the nominal $G_{opt}$ obtained after post-layout simulations throughout our measurements.

\begin{table}[]
\setlength\tabcolsep{1.5pt}
    \centering
    \caption{Range of Optimal NCF Gain $G_{opt}$. A Commercial Temperature Range From 0$^{\circ}$C to 70$^{\circ}$C was Considered.}
    \label{tab:G_range}
    \begin{tabular}{|c|c|c|c|c|c|c|c|c|c|}
    \hline
    Temp    &\textbf{0$^{\circ}$C}    &70$^{\circ}$C   &0$^{\circ}$C    &70$^{\circ}$C   &27$^{\circ}$C   &0$^{\circ}$C    &70$^{\circ}$C   &0$^{\circ}$C    & \textbf{70$^{\circ}$C}\\ \hline
    Process &\textbf{SS}     &SS     &SF     & SF    &TT     &FS     &FS     &FF     & \textbf{FF}\\ \hline
    $G_{opt}$&\textbf{ 1.27} &1.24   &1.16   &1.13   &1.11   &1.10    &1.10   &0.98   & \textbf{0.98} \\ \hline
    Mismatch &\textbf{13}    &10     &3      &1      &-      &-2   &-2   &-13   &\textbf{-13} \\
    w.r.t. nominal [\%] & & & & & &  & & &\\ \hline
    \end{tabular}
\end{table}
      
\section{Prototype Measurements\label{sect:meas}}
To demonstrate the performance of our architecture, a prototype was manufactured in a 28nm CMOS technology node. Multiple supply voltage domains are present which can be configured individually. More specifically, the supply voltages of the first and second VCOs, the sense amplifiers \& XOR gates and the digital core can all be set separately. The sampling frequency was set at 3.5 GHz. The supply voltages of the digital core and the sense amplifiers were set at 1.05 V and 1 V respectively. All other voltages were set at the nominal supply voltage of 0.9 V.

\begin{figure}
    \centering    
    \begin{minipage}{0.45\columnwidth}
    \flushleft{(a)} \\
    \includegraphics[width=0.9\columnwidth, angle =90 ]{{Images/20a.pdf}} 
    \end{minipage}
    \begin{minipage}{0.45\columnwidth}
    \flushleft{(b)}\\
    \includegraphics[width=0.9\columnwidth]{{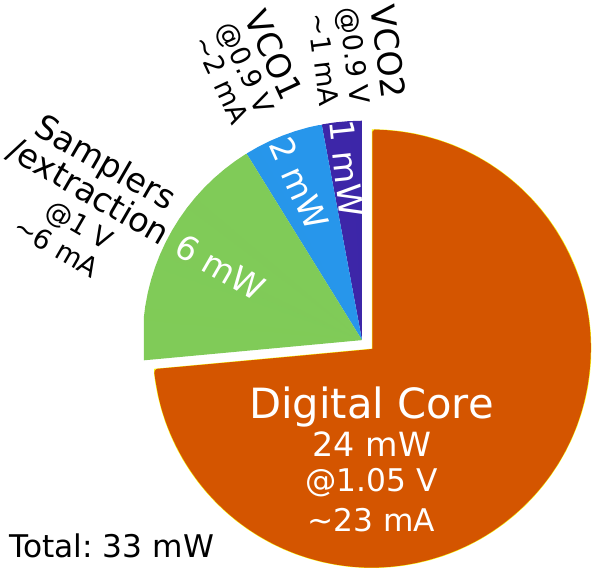}}
    \end{minipage}
    \begin{minipage}{0.45\columnwidth}
    \flushleft{(c)}\\
    \includegraphics[width=0.9\columnwidth]{{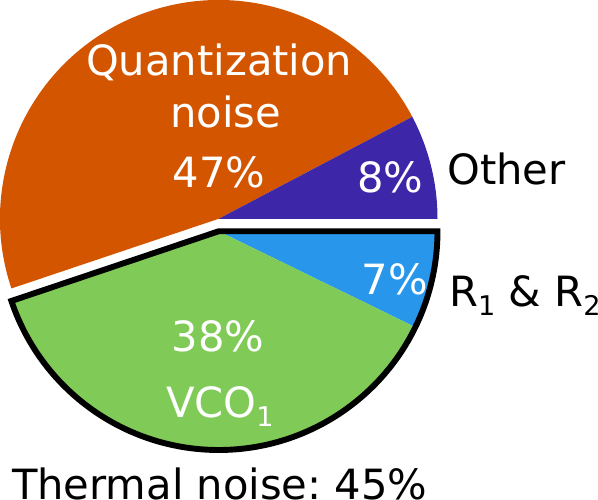}}
    \end{minipage}
    \caption{Chip micrograph (a), power and current consumption for a full-scale input signal (b) and noise breakdown (c).}\label{fig:chip}
\end{figure}
A micrograph of the chip is shown in Fig.~\ref{fig:chip}(a).
The total core area is only $0.015\,\mathrm{mm}^2$, including $0.006\,\mathrm{mm}^2$ for the analog core. The circuit consumes $9 \ \mathrm{mW}$ and $24 \ \mathrm{mW}$ in the analog and digital core respectively for a full-scale input signal of $\mathrm{900\ mV_{pp}}$ as summarized in Fig.~\ref{fig:chip}(b). Fig.~\ref{fig:chip}(c) shows a breakdown of the dominant noise sources.

\subsection{Measured Nonlinearity Curve and Calibration Procedure\label{subsect:cal}}
\begin{figure}[t]
    \includegraphics[width=1\columnwidth]{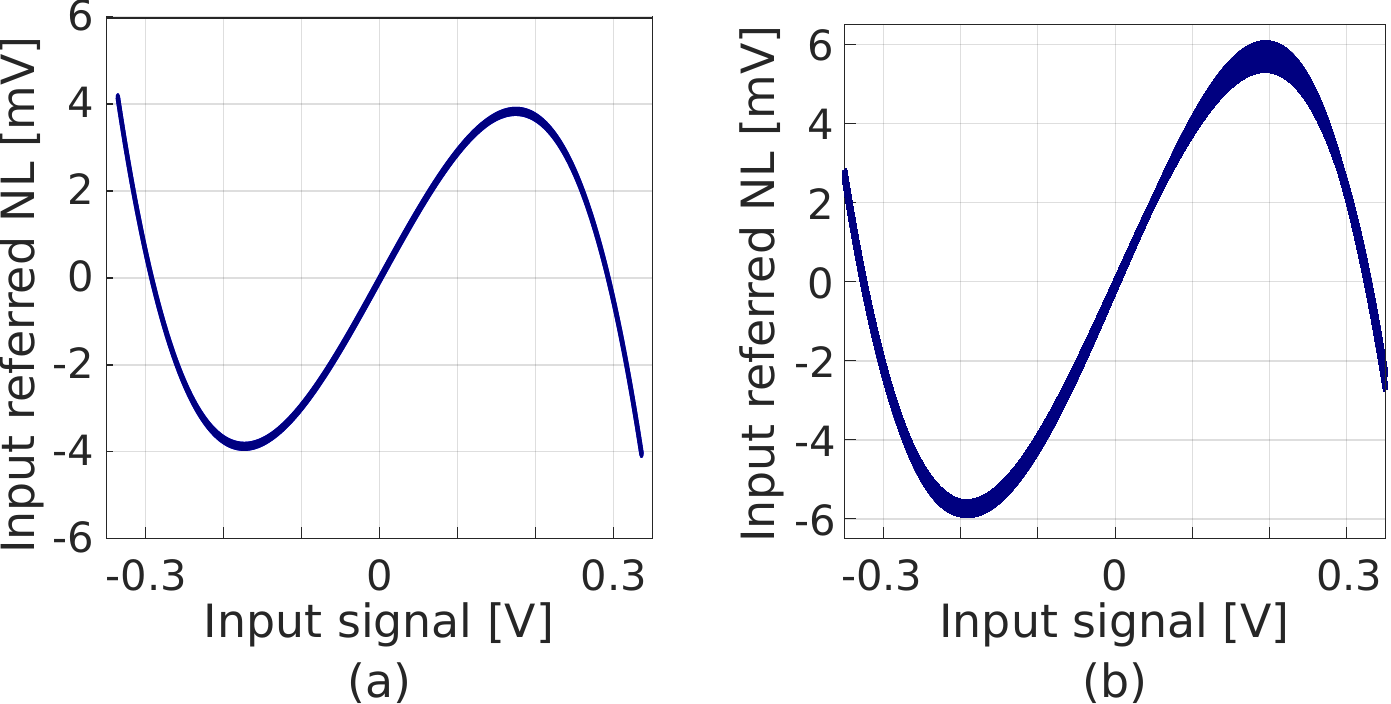}
    \caption{{{Measured distortion for a 2 MHz (a) and 26.5 MHz (b) $\mathrm{750\ mV_{pp}}$ sinusoidal input signal.}}}\label{fig:NL}
\end{figure}
\begin{figure}[t]
    \center
    \includegraphics[width=\columnwidth]{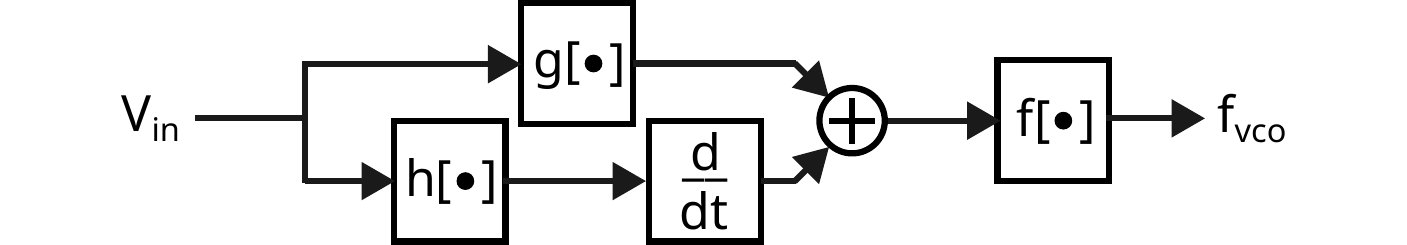}
    \caption{Model of the VCO using the resistive driver.} \label{fig:model_vco}
\end{figure} 
\begin{figure}[t]
    \center
    \includegraphics[width=\columnwidth]{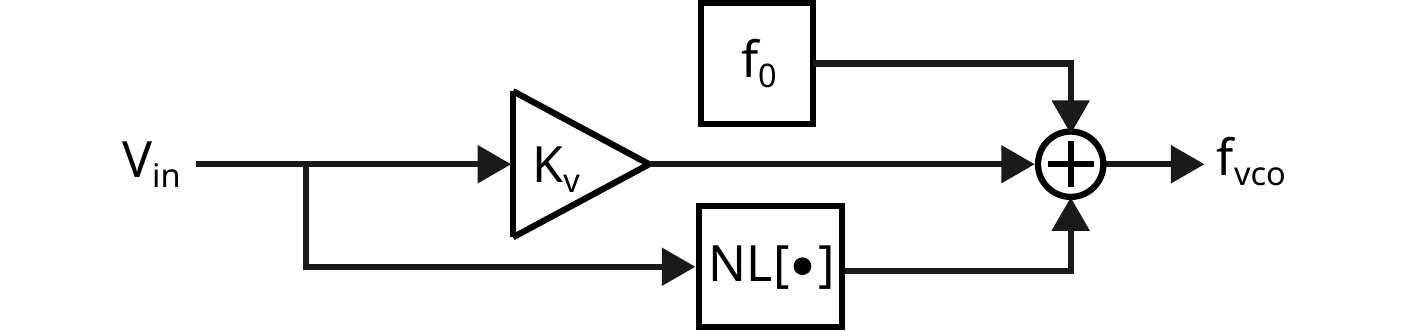}
    \caption{Redrawn model of Fig.~\ref{fig:model_vco}.} \label{fig:redrawn}
\end{figure} 
\begin{figure}[t]
    \center
    \includegraphics[width=\columnwidth]{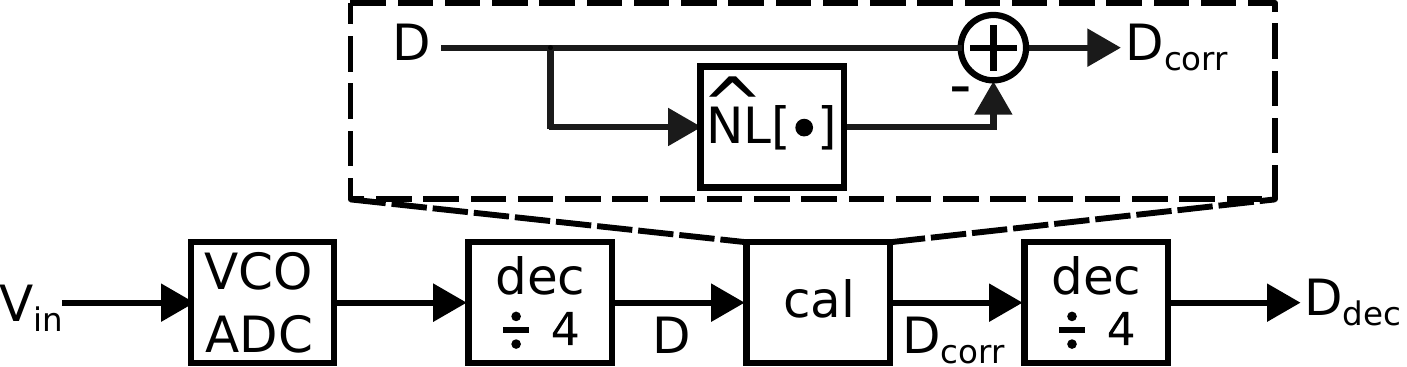}
    \caption{Block diagram of the calibration procedure, with non-linearity correction calibration ('cal') and decimation steps including prior decimation filter ('dec').} \label{fig:cal}
\end{figure}  
In a first step, we will evaluate the distortion of the ADC. For this, we perform calibration measurements where the ADC is driven by a spectrally pure sine wave and the output $D$ is captured. The output can then be split into three components: (i) the fundamental sine wave, which corresponds to the signal component $D_\textrm{sig}$, (ii) the distortion $\textrm{DIST}$, which corresponds to the distortion of $D_\textrm{sig}$ and (iii) the noise $r$. This is done using a fit of the components at the input frequency $f_{in}$ and harmonics of $f_{in}$, leading to $D_{sig}$ and DIST. The residue is the noise $r$. For this procedure, efficient sine-wave fitting algorithms as developed in IEEE-STD-1057 can be used \cite{saux_2024}. For future reference, we introduce the clean output signal, which is the output signal without the noise: $D_{cl}[n]=D[n]-r[n]$. Here $n$ represents the n-th sample. 

{The extracted signal component $D_{sig}$ and distortion DIST can be referred to the input to obtain what we will call the instantaneous distortion, ie.~the distortion with respect to the input signal for every sample. The resulting curves for a 2 MHz and 26.5 MHz signal are shown in Fig.~\ref{fig:NL}(a) and \ref{fig:NL}(b) respectively.} A clear frequency-dependent effect can be observed, as the distortion noticeably increases for higher frequencies.
This can be explained by parasitic capacitances at the tune node of the VCO influencing the VCO current and frequency \cite{saux_2024}. Traditional VCO ADC calibration methods as used in \cite{amir_2017, wu202016, baert2020} model the VCO nonlinearity by a single nonlinear function and are not capable of correcting the frequency-dependent distortion components.

{Instead, the linear performance of the ADC is improved by using the foreground calibration procedure we recently published in \cite{saux_2024}, which explicitly takes into account these frequency-dependent components. We will provide an overview of the applied procedure in this section and refer to the aforementioned work for an in-depth discussion.}
It is shown in \cite{saux_2024} that a VCO can be modeled as in Fig.~\ref{fig:model_vco}, where $g[\cdot]$, $h[\cdot]$ and $f[\cdot]$ are nonlinear functions. The nonlinear components can be isolated from the linear and constant components, ie.~the free-running frequency and frequency gain, into an isolated distortion function $NL[\cdot]$. This results in the redrawn Fig.~\ref{fig:redrawn}. An estimate $\widehat{NL}[\cdot]$ of $NL[\cdot]$ can now be written in function of $D_{cl}$ as:
\begin{equation}
    \widehat{NL}(D_{cl}[n])\approx \sum_{k=1}^{N_k} \hat{c}_{k}\left[ \sum_{i=0}^{N_i} \hat{a}_i D_{cl}^i[n] + \sum_{j=1}^{N_j} \hat{b}_j \Delta D_{cl}^j[n] \right]^k 
    \label{eq:nl}
\end{equation}
with coefficients $\hat{a}_i$, $\hat{b}_{j}$ and $\hat{c}_k$. The difference $\Delta D_{cl}^j[n] = (D_{cl}^j[n]-D_{cl}^j[n-1])$ is used to approximate the time derivative originating from the capacitive effects mentioned above.  $N_i$, $N_j$, $N_k$ represent the highest powers of the polynomials used for the approximation. During the fitting step of the calibration, the coefficients of $\widehat{NL}[\cdot]$ are determined by fitting  $\widehat{NL}[\cdot]$ to DIST using least square minimizations.

Once the estimated coefficients $\hat{a}_i$, $\hat{b}_{j}$ and $\hat{c}_k$ are known, these can be used for correction of the ADC output. During normal operation, $\widehat{NL}(D)$ is subtracted from the ADC output to obtain the corrected output $D_{corr}$, as shown in Fig.~\ref{fig:cal}. Also, as first described in \cite{amir_2017}, the nonlinearity correction is executed on a partially decimated signal including a prior anti-aliasing filtering. This improves performance by removing out of-band shaped quantization noise that would otherwise be partly converted to in-band white noise after passing through the nonlinearity correction block. 

%remark
Note that despite the 1-1 MASH being more complex than the single-stage VCO ADC calibrated in \cite{saux_2024}, it is expected the same procedure can still be applied. The first stage will be the dominant source of distortion of $V_{in}$ and can be modeled as Fig.~\ref{fig:model_vco}, as it is simply a single-stage VCO. While the second stage will introduce distortion of $E$, no significant distortion of the global input signal $V_{in}$ due to the second stage is expected. This is a consequence of (2), where it is shown that $E$ is an approximation of the quantization error of the first stage. As mentioned before, the nonlinearity in the second  stage is mainly of concern for the $Q_1$ noise leakage, which has little to no input signal dependency. Note that this cannot be corrected by the calibration procedure, which is one of the reasons the cross-coupling technique was introduced in section 3.

In present work, the calibration is performed using $N_i=N_j=5$ and $N_k=2$.
The nonlinearity correction itself is implemented off-chip in a bit-accurate manner. $D$ and $D_{corr}$ are represented as 14-bit fixed point numbers. Other calculations are performed with the same accuracy. The polynomials $\sum_{i=0}^{N_i} \hat{a}_i D^i[n]$ and $\sum_{j=1}^{N_j} \hat{b}_j D^j[n]$ of $\widehat{NL}[\cdot]$ are approximated using 512-element LUTs containing the polynomial values and linear interpolation is performed between the 2 points closest to the input of the correction block. This is similar to the method first developed in \cite{baert2020}. To evaluate the cost of an on-chip implementation, an implementation was generated using automatic synthesis and place and route tools. The estimated additional power consumption and area for the correction would be 2.2 mW (of which 1.4 mW in the LUTs) and 0.0015 $\mathrm{mm^2}$ respectively. The results of this calibration procedure will be discussed in the next section.
\vspace{-1em}
\subsection{Single- and Two-tone Measurement Results}
\begin{figure}[t!]
    \centering    
    \includegraphics[width=1\columnwidth]{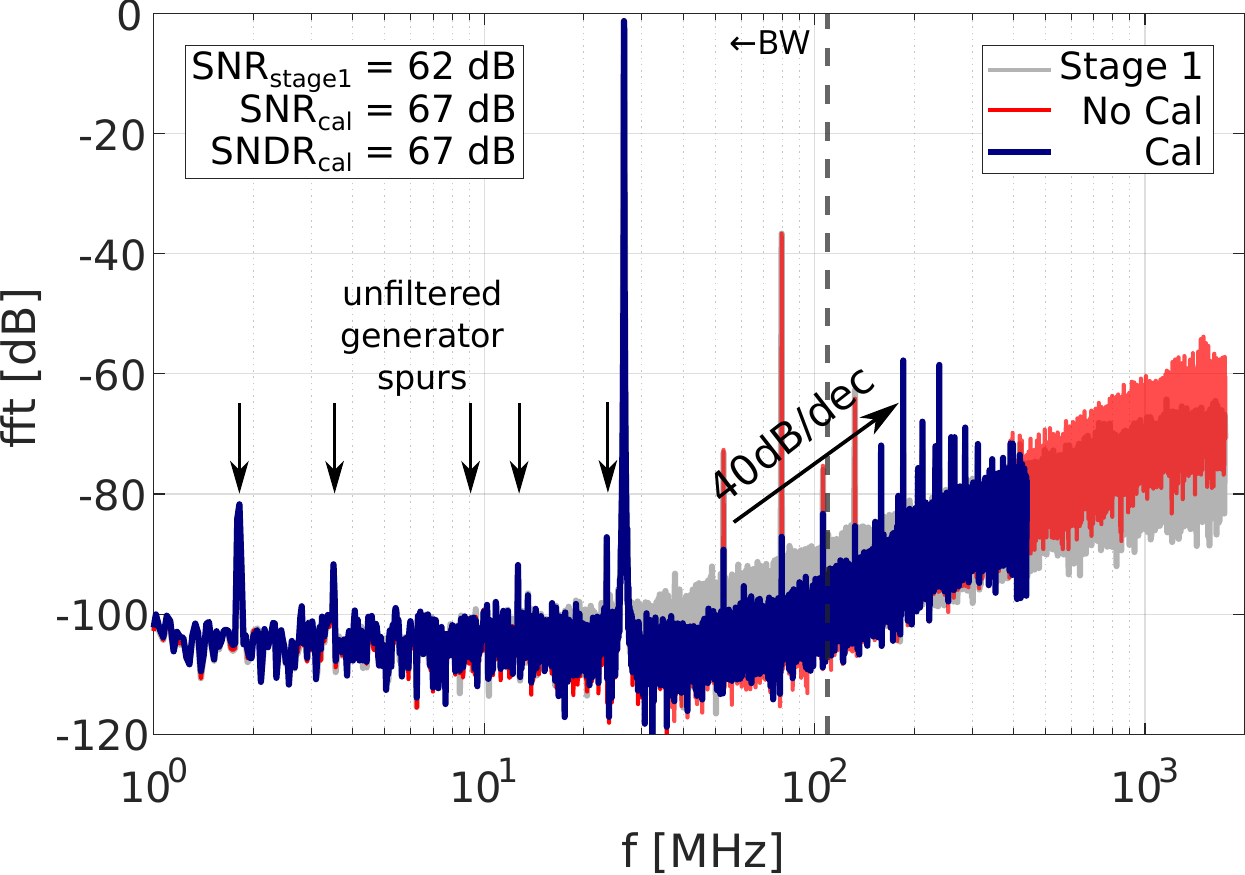}
    \caption{{131K-point FFT averaged three times for a $\mathrm{26.5\ MHz\ 750\ mV_{pp}}$ input.}}
    \label{fig:fft1s}
\end{figure}

Fig.~\ref{fig:fft1s} shows the output frequency spectrum for a 26.5 MHz $\mathrm{750\ mV_{pp}}$ sinusoidal input. The input signal was filtered by a 27 MHz low-pass filter.  Some lower-frequency generator spurs are therefore present in the output spectrum, but do not affect the measurement interpretation.
A bandwidth of $109.375\,$MHz ($\mathrm{OSR}=16$) is used.
The gray curve shows the output frequency spectrum when only reading out the first stage. This shows a significant amount of in-band first-order shaped noise, resulting in $\mathrm{SNR}=62\,$dB.
For the 1-1 MASH (red curve), an SNR of $67 \,$dB and uncalibrated SNDR of $36\,$dB are obtained. This uncalibrated SNDR is largely in line with the design target of section \ref{subsect:ring_oscillator} to limit the nonlinearity of the first stage to about 1\%.
Applying the calibration procedure described above to the 1-1 MASH leads to the blue curve and an improved $\mathrm{SNDR}=67\,$dB.

\begin{figure}
\vspace{-1em}
    \centering        \includegraphics[width=1\columnwidth]{{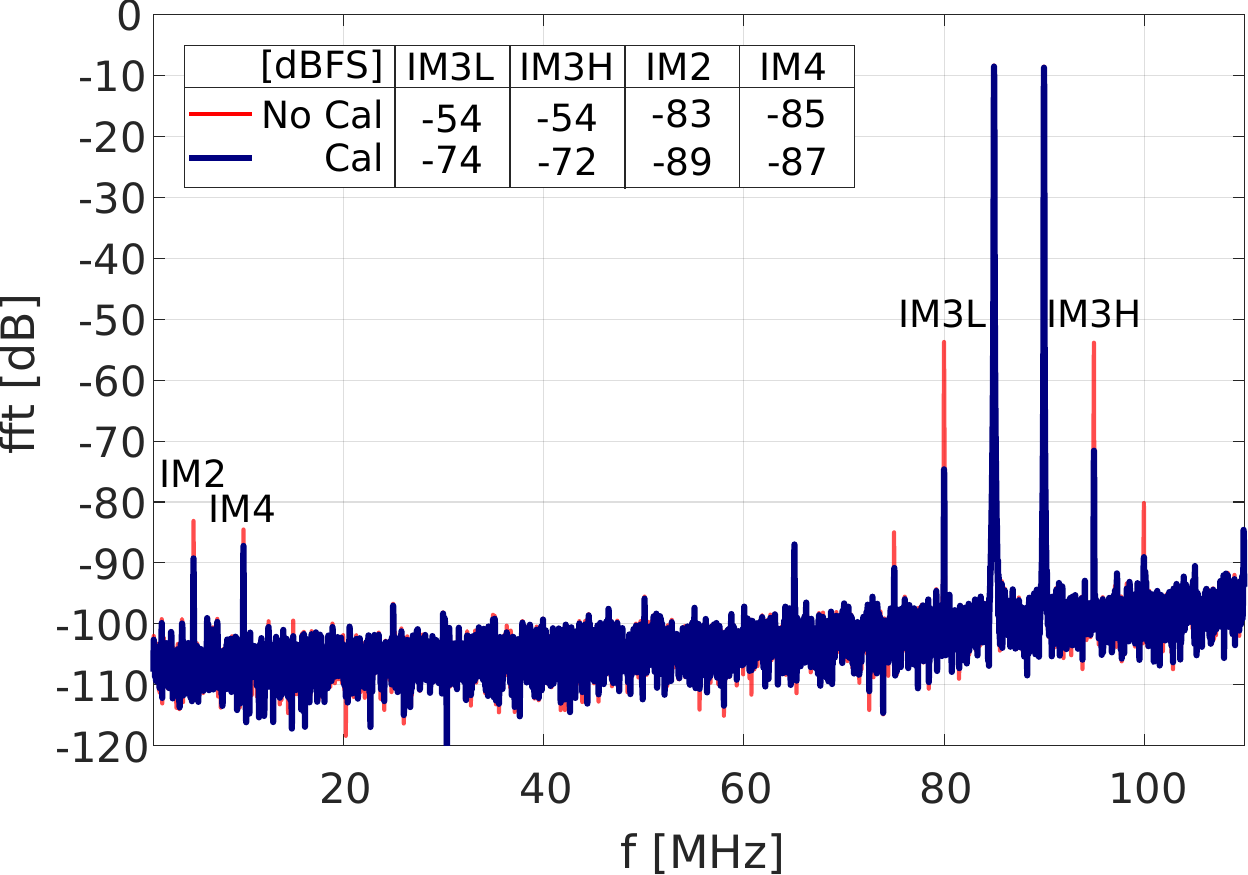}} 
    \caption{131K-point FFT averaged three times for a twotone measurement with 85 and 89.99 MHz -8.5 dBFS inputs.} 
    \label{fig:fft_twotones}
\end{figure}

Fig.~\ref{fig:fft_twotones} shows the result of a two-tone test with 85 and 89.99 MHz -8.5 dBFS input signals, where all harmonics are under $-70$ dBFS after calibration. 

\begin{figure}[h!]
\vspace{-1.1em}
    \centering
    \includegraphics[width=1\columnwidth]{{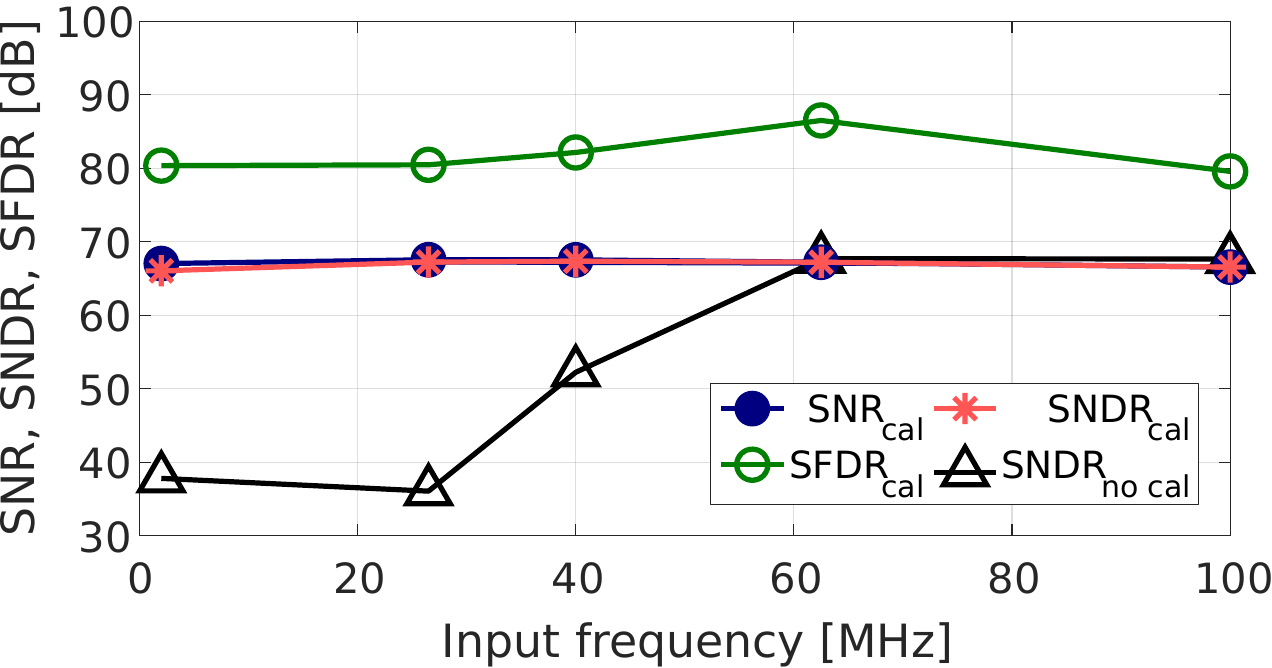}}
    \caption{{SNDR, SNR and SFDR in function of the frequency of a $\mathrm{ 750\ mV_{pp}}$ sine wave.} }\label{fig:SNR_FREQ}
\end{figure}

{The improvement after calibration is also visible in Fig.~\ref{fig:SNR_FREQ}, which shows the achieved SNDR before calibration and the SNR, SNDR and SFDR after calibration for a $\mathrm{ 750\ mV_{pp}}$ input for a range of input frequencies. The SFDR after calibration is over 80 dB for all measured frequencies. }

\begin{figure}
\vspace{-1.5em}
    \centering	\includegraphics[width=1\columnwidth]{{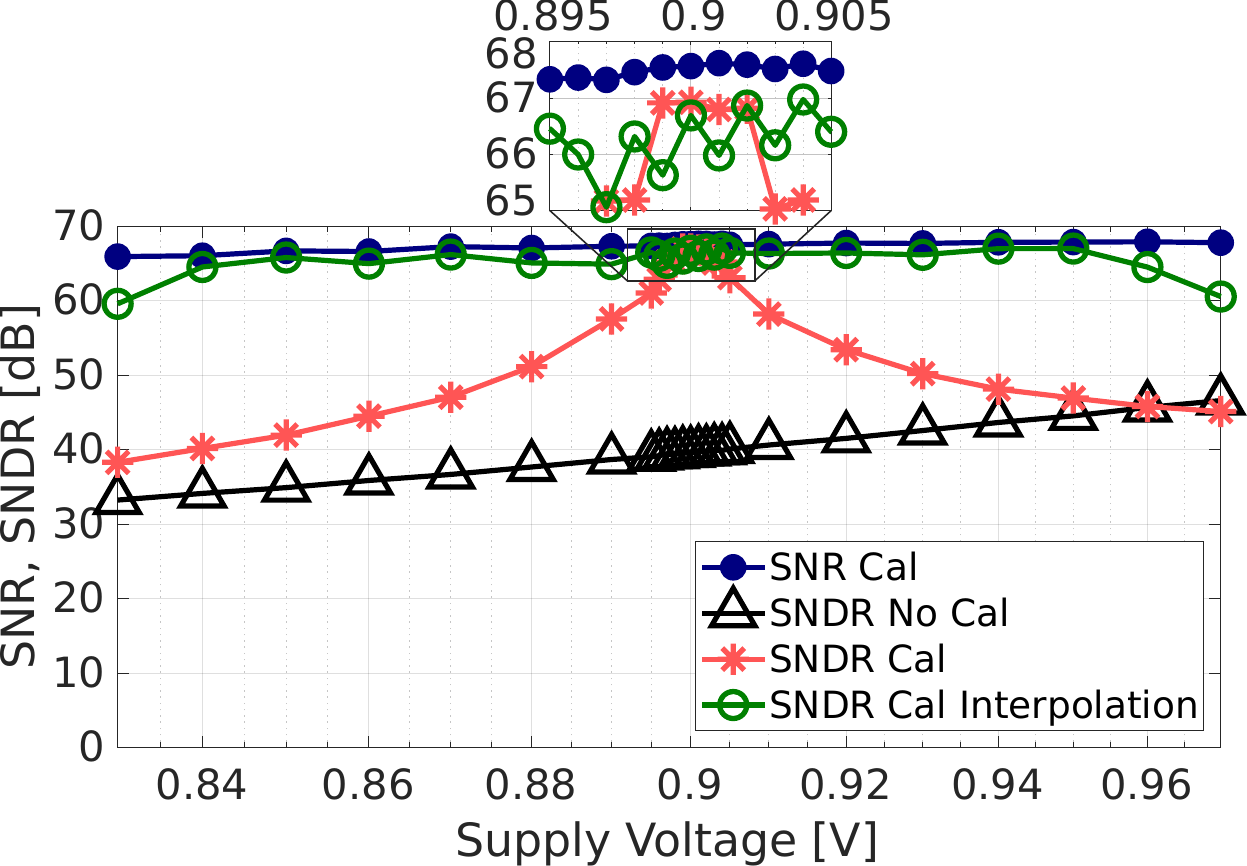}}
    \caption{SNR and SNDR for different VCO supply voltages.} \label{fig:VDD_sweep}
\end{figure}

Fig.~\ref{fig:VDD_sweep} illustrates the influence of the VCO supply voltage on the obtained SNDR and SNR. Due to the PVT dependency of VCO nonlinearity, using a foreground calibration procedure (commonly used in most high-bandwidth VCO ADCs \cite{gielen2020time2, wu2019, baert2020}) results in a calibration performance that varies with changes in temperature and voltage. For example, the results shown in Fig.~\ref{fig:VDD_sweep} closely align with those of \cite{baert2020}.  Fig.~\ref{fig:VDD_sweep} shows a narrow range of around 7 mV where the calibrated SNDR (red curve) remains above 65 dB, after which it drops. 

To address this, one solution is recalibrating the ADC when the supply voltage changes to maintain a nearly constant SNDR \cite{baert2020}, although this may not always be practical. Another approach involves interpolation. It is possible to perform calibration measurements over multiple supply voltages at calibration time. As many modern SoCs are equipped with PVT monitors, an interpolation can be performed over the obtained calibration coefficients during regular operation using the information from the on-board voltage sensor. This approach is illustrated by the green curve of Fig.~\ref{fig:VDD_sweep}, where quadratic interpolation between calibration coefficients obtained at 0.85 V, 0.9 V and 0.95 V was performed, achieving an SNDR of 65 dB over a 110 mV range between 0.84 V and 0.95 V. An error of 1.92 mV was introduced in the supply voltage estimate used for the interpolation, to mimic the limited resolution of published voltage sensors in our technology node \cite{ du_2023}. Finally, it is also common practice in many high-performance chips to use voltage regulators such as a low-dropout regulator (LDO) to isolate sensitive ADCs and analog components from the noisy supply of digital circuits \cite{razavi_2019, Javvaji_2024}. Many recently published LDOs offer line regulations that keep variations well within the previously mentioned 7 mV range \cite{yang_2024}, therefore also effectively addressing this issue. The interpolation method using a temperature sensor can also be employed to address the effects of temperature fluctuations.

\begin{figure}[h!]
\vspace{-1.5em}
    \centering
    \includegraphics[width=1\columnwidth]{{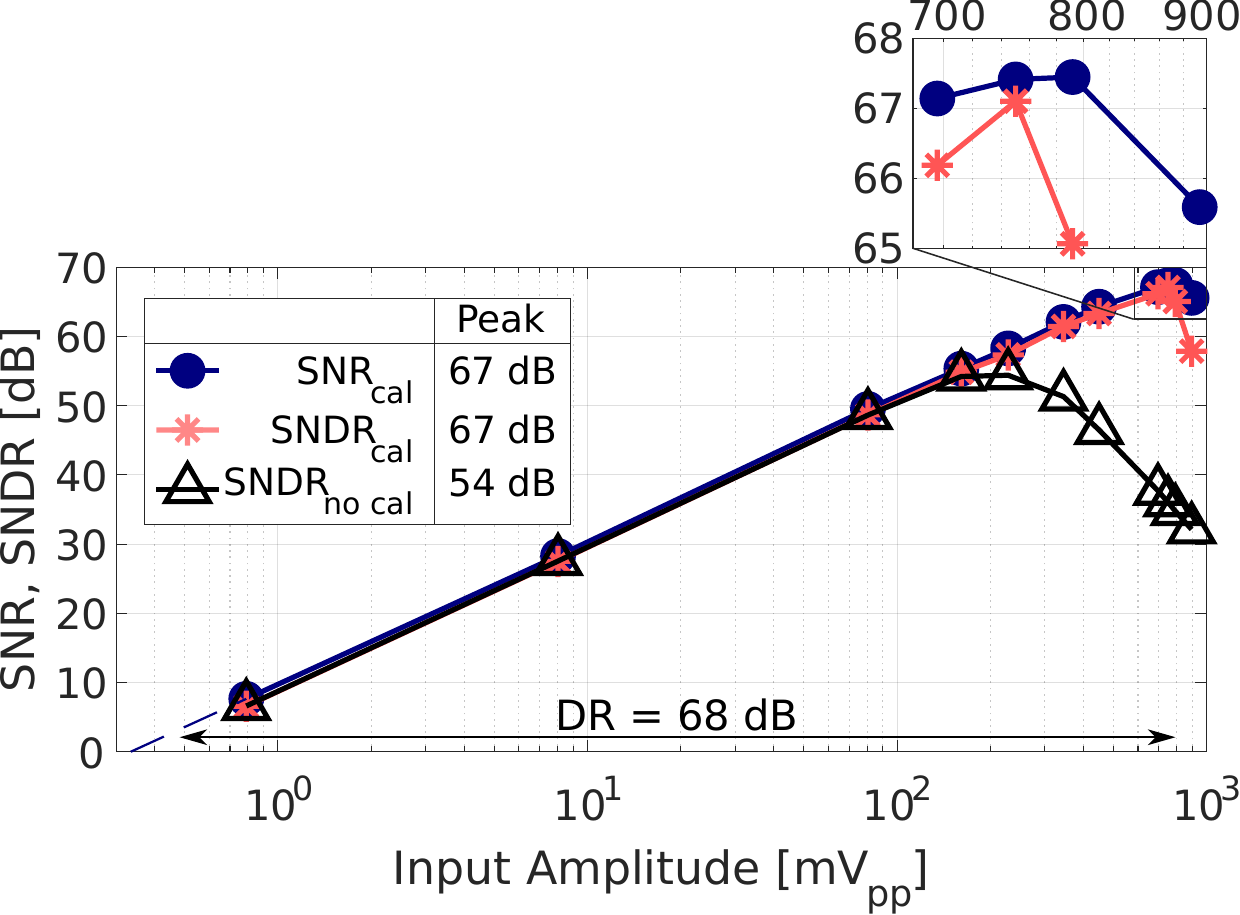}}
    \caption{{SNR and SNDR in function of the amplitude of a 26.5 MHz sine wave.}} \label{fig:DR_SNDRs}
\end{figure}

Fig.~\ref{fig:DR_SNDRs} shows the SNDR in function of the input signal amplitude of a 26.5 MHz sine before and after calibration. We obtain a dynamic range $\mathrm{DR}=68\,$dB. We also find a peak $\mathrm{SNDR}=67\,$dB for a $\mathrm{ 750\ mV_{pp}}$ signal at 26.5 MHz. Dynamic range measurements on an extra chip from the same wafer resulted in similar results to those reported in this work.

Note that the same set of calibration coefficients, which were derived from an initial calibration measurement, were used for all further measurements in this work (except for the green curve in Fig.~\ref{fig:VDD_sweep}). These measurements clearly support our claim that the 1-1 MASH can be calibrated in the same way as a single-stage VCO ADC.
\begin{table*}
    \centering
    \begin{threeparttable}
    \caption{
    Comparison With State-of-the-art VCO-only ADCs \cite{wu2019, wu202016, xing201542, amir_2017} in Order of Bandwidth, a Pipelined ADC With a VCO-based Second Stage \cite{shibata202016} and a Recently Published Conventional $\Sigma \Delta$ Modulator \cite{Javvaji_2024}.}\label{table:SOTA}
    {\small
    \begin{tabular}{|c||c||c|c|c|c|c|c|}
    \hline
                                                    
                                   & \textbf{This}    & \cite{shibata202016} & \cite{wu2019}    & \cite{wu202016}   &  \cite{xing201542} &  \cite{amir_2017} & \cite{Javvaji_2024}      \\
                                   & \textbf{Work}    &   ISSCC              & JSSC             & ISSCC             &   JSSC             &   JSSC            & JSSC                     \\
                                   &                  &   2020               & 2019             & 2020              &   2015             &   2017            & 2024                     \\\hline  \hline
                                                                                                                                        
    Technology Node   [nm]         & 28 	          & 16                   & 65               & 28                & 40                 &  65               & 28                       \\ \hline
    OTA-free                       & Yes              & No                   & Yes              & Yes               & Yes                & Yes               & No                       \\ \hline
    Type                           & 1-1 MASH         & Pipeline +           & NUS              & NUS               &2-step              &$\Sigma \Delta$    & $\Sigma \Delta$          \\
                                   &  VCO             & VCO                  & VCO              & VCO               &VCO                 & VCO               &    Conventional          \\ \hline
    Noise Transfer Function Order  & 2                & 1                    & 1                & 1                 & 1                  & 3                 & 4                        \\ \hline
    Sampling 	Rate [GS/s]        &  3.5 	          & 6.4                  & 4                & 2                 & 1.6                & 1.6               & 6                        \\ \hline
    Bandwidth [MHz]				   &  109.375 	      & 800                  & 200              & 40                & 40                 & 10                & 120                      \\ \hline
    Power [mW]       		       &  33 + 3.7*       & 281                  & 49.7             & 55.5              & 2.57               & 3.7               & 115.0                    \\ \hline
    SNDR [dB]	         	       & 67	              &  58                  & 57               & 76                & 60                 & 66                & 72.8                     \\ \hline
    DR [dB]                        & 68               & 60                   & 60               & 78                & 62                 & 71                & 73.4                     \\ \hline \hline
    $\mathrm{FOM}_\mathrm{S}$ [dB]  & 162*            & 153                  & 153              & 165               & 161                & 160               & 163                      \\ \hline
    $\mathrm{FOM}_\mathrm{DR}$ [dB] & 163*            & 155                  & 156              & 167               & 164                & 165               & 164                      \\ \hline
    Core Area [$\mathrm{mm}^2$]     &  0.015 / 0.017* &0.34                  & 0.13             & 0.023             & 0.017              & 0.01              & 0.146                    \\ \hline
    \end{tabular}                                                                                                                        
  {\footnotesize
    \begin{tablenotes}[para]
    \item[*] Including estimated power and area of nonlinearity correction and NCF \\
    \item[a] NUS = non-uniform sampling
    \item[b] $\mathrm{FOM}_\mathrm{S}=\mathrm{SNDR} + \mathrm{10log_{10}(BW/Power)}$
    \item[c]$\mathrm{FOM}_\mathrm{DR}=\mathrm{DR} + \mathrm{10log_{10}(BW/Power)}$
    \end{tablenotes}
}
}
\end{threeparttable}
\end{table*}
Based on these measurements, Table \ref{table:SOTA} summarizes the performance of our prototype and compares it to state-of-the-art VCO ADCs \cite{shibata202016, wu2019, wu202016, xing201542, amir_2017} and a recent $\Sigma \Delta$ modulator \cite{Javvaji_2024}. From the table it is clear that our prototype compares favorably with these designs, achieving an excellent bandwidth, Figure-of-Merit~(FoM) and area.



\section{Conclusion\label{sect:conc}}
In this work, an open-loop 1-1 MASH VCO ADC architecture was presented and a prototype was manufactured in 28nm CMOS. By using the estimated quantization noise of a first VCO ADC stage as the input of a second stage and combining both outputs using the correct noise correction filters, second-order noise shaping was achieved. To the authors' knowledge, this results in the first higher-order purely VCO-based ADC operating at more than 100 MHz bandwidth. A key enabler is the use of error estimation on all phases of the first stage, which allows to significantly reduce the in-band PFM spurs that limit the encoding accuracy. The ADC core consumes $33\,$mW.
An SNDR of $67\,$dB and a DR of $68\,$dB are obtained with a bandwidth of $109.375\,$MHz. Including the estimated power and area of off-chip components, this results in a $\mathrm{FoM_{DR} = 163}$ dB and core area of $\mathrm{0.017\,\mathrm{mm}^2}$.
%\clearpage
%\newpage
\bibliographystyle{IEEEtran}
\end{document}